\documentclass[final,5p,times,twocolumn]{elsarticle}

\usepackage{amsmath}
\usepackage{xcolor}
\usepackage{xspace}
\usepackage{amsthm}
\theoremstyle{definition}
\newtheorem{dfn}{Definition}

\usepackage{graphicx}
\graphicspath{{figs}}

\usepackage{multicol}
\usepackage{multirow}

\usepackage{balance}

\input{defs.sty}

\journal{Journal of Systems and Software}
\date{}

\begin{document}

\begin{frontmatter}

\title{An Empirical Study on the Impact of Change Granularity in Refactoring Detection}

\author[inst1]{Lei Chen}
\affiliation[inst1]{organization={School of Computing, Institute of Science Tokyo},
            addressline={Ookayama 2--12--1, Meguro-ku}, 
            city={Tokyo},
            postcode={152--8550},
            country={Japan}}
\author[inst1]{Shinpei Hayashi}

\begin{abstract}
Detecting refactorings in commit history is essential to improve comprehension to code changes on code reviews, and to provide valuable information for empirical studies on software evolution.
Techniques have been proposed to accurately detect refactorings on the granularity of a single commit.
However, refactorings can be made over multiple commits because of their complexity or other practical development problems, which cause detecting on only the granularity of a single commit not enough.
We observe that some refactorings can only be detected in coarser granularity, i.e., changes conducted over multiple commits, or in the granularity of a single commit but not in coarse-grained.
We call these types of refactorings as \emph{coarse-grained refactorings}~(CGRs) and \emph{ephemeral refactorings}~(EPRs).
We investigated the features and causes of CGRs and EPRs through an empirical study of 32 open-source Java projects and found that both commonly occur during development.
In addition, we found that refactoring types related to splitting or merging classes and packages, as well as those involving modifications to the inheritance structure, tend to be CGRs, and types targeting small objects such as variables and attributes, and refactorings with context-sensitive detection criteria tend to be EPRs.
The causes of CGRs and EPRs are analyzed and categorized, and the relationships between the commit messages of CGRs and themselves are also assessed.
We found that about 20\% of commit messages explicitly suggest the existence of CGRs.
We suggest that CGRs and EPRs be valued in refactoring research and that detectors be extended to identify CGRs.
\end{abstract}


\end{frontmatter}

\section{Introduction}\label{sec:introduction}

Refactoring mining within the commit history benefits both software developers and researchers.
Refactoring is the process of improving the internal structure of code without changing its external behavior~\cite{fowler2018refactoring}.
For developers, it facilitates the understanding of code changes~\cite{tsantalis2018accurate} and enhance the effectiveness of code reviews~\cite{kim2012field}.
In terms of researchers, it can provide valuable information for empirical studies on software evolution, such as the benefits of refactorings~\cite{kim2014empirical}, the relationship between bugs and refactorings~\cite{bavota2012does}.

To mine refactorings, recent studies proposed refactoring detectors that detect refactorings by comparing two source code snapshots~\cite{tsantalis2018accurate,dig2006automated, kim2010ref, prete2010template, weissgerber2006identifying, silva2017refdiff, silva2020refdiff,Tsantalis:TSE:2020:RefactoringMiner2.0}.
While traditional approaches focus on detecting refactorings between releases~\cite{dig2006automated, kim2010ref, prete2010template,weissgerber2006identifying}, modern detectors such as RefDiff~\cite{silva2017refdiff,silva2020refdiff} and RefactoringMiner~\cite{tsantalis2018accurate,Tsantalis:TSE:2020:RefactoringMiner2.0} analyze individual commits, which means that two snapshots before and after a single commit are compared.
These methods have achieved high accuracy in detecting refactoring within commits~\cite{ref-det-tools}.

However, refactorings may be performed over multiple commits.
For example, when moving a method from one class to another, some developers choose to first copy the implementation of the method to the target-of-move class in the first commit.
After confirming that users will not be affected, the original implementation is removed in the next commit.
In this case, the \MoveMethod refactoring could not be detected from either of the two commits using commit-level refactoring detectors, but it becomes detectable from a coarse-grained granularity that spans across both commits.

By investigating the impact of changing the refactoring detection granularity from a single commit to coarser-grained, we are able to gain a deeper insight into the refactorings similar to the above.
We define a refactoring whose code changes are recorded across multiple commits, rather than confined in a single commit as \emph{coarse-grained refactoring}~(CGR).
Conversely, a refactoring whose code changes are recorded within a single commit and where modifications in adjacent commits would disrupt their integrity are referred to as \emph{ephemeral refactoring}~(EPR).
We use the term \textit{ephemeral} to emphasize its brief lifespan, as these refactorings are confined to a narrow historical scope of commits and will disappear in a broader scope.
An example of EPR is that a refactoring is initially applied and then reverted due to being considered an inappropriate change.

The existence of CGRs and EPRs suggests that relying solely on a single specific detection granularity is not enough, as CGRs may be overlooked, and EPRs may require analysis across multiple commits to be detected. This limitation may lead to incomplete or incorrect conclusions in empirical studies, such as refactoring type and software evolution analysis.

Also, uncovering those refactorings can help developers gain a better understanding of code evolution.
To investigate the features and causes of CGRs and EPRs, we conduct an empirical study on open-source Java repositories.

We regard the original commits recorded in the commit history as fine-grained and name them fine-grained commits~(FGCs). 
To change the granularity of commits, multiple FGCs are squashed into one to form a coarse-grained commit~(CGC).
In this way, the code changes from multiple FGCs are combined into a CGC.
The number of FGCs squashed into one CGC is referred to as \textit{granularity level}.
Refactoring detection is conducted on both FGCs and CGCs using the state-of-the-art tool RefactoringMiner~\cite{tsantalis2018accurate,Tsantalis:TSE:2020:RefactoringMiner2.0} to extract CGRs and EPRs for analysis.

The main contributions of this paper are shown below:
\begin{itemize}
    \item We defined the concept of CGR and EPR and proposed a technique to detect them from the commit history.
    \item An empirical study is conducted on a dataset of 32 open-source Java repositories.
    \item The features and causes of CGRs and EPRs are analyzed.
    \item We investigated the relationship between developers' intentions deduced from commit messages and CGRs.
    We discussed the implications of CGRs and EPRs for researchers and practitioners.
\end{itemize}

The findings of this paper are as follows:
\begin{itemize}
    \item CGRs and EPRs are common in development, and on average, per 100 refactorings conducted, there are also 5.71 CGRs and 9.89 EPRs be performed.
    \item Refactoring types related to splitting or merging classes and packages, as well as those involving modifications to the inheritance structure, tend to be CGRs while types related to operations on small objects such as variables, attributes, and parameters tend to be associated with EPRs.
    \item The cause of CGR can be categorized into two types: \textit{Generation} and \textit{Combination} according to their composition.
    While the cause of EPR can be divided as \textit{Disruption}, \textit{Absorption}, and \textit{Covered}.
    \item We found that 20\% commit messages explicitly express the existence of CGRs, and the characteristics of those commit messages are summarized and categorized into four types.
\end{itemize}

This paper is an extension of work reported originally in the proceedings of the 30th IEEE/ACM International Conference on Program Comprehension~\cite{chen2022impact}.
The main improvements are the following:
\begin{itemize}
    \item A more delicate matching scheme is designed to mine more previously undiscovered CGRs.
    \item Proposed the definition of EPRs and also the technique to mine them.
    \item Conducted an experiment on a larger-scale dataset that contains 32 open-source Java repositories.
    \item Analyzed the features from more aspects for CGRs and the features for EPRs.
    \item Investigate the causes of EPRs.
    \item Implications for both researchers and practitioners.
\end{itemize}

The remainder of this paper is organized as follows.
\Cref{sec: coarse-grained_and_fine-grained_refactorings} explains the definition of CGR and EPR.
\Cref{sec:related work} introduces the background of this study.
In \cref{sec: detecting coarse-grained refactorings and ephemeral refactorings}, we explain the matching mechanism of refactorings at different granularities that we designed.
\Cref{sec:methodology} introduces the overview of our study.
We extract CGRs and EPRs on the dataset and analyze and discuss five research questions in \cref{sec: empirical_study}.
The possible threats to validity are discussed in \cref{sec:threats_to_validity}, and the implications are discussed in \cref{s: Implication}.
Finally, we conclude our work and state our plans for future work in \cref{sec:conclusion_and_future_work}.

\section{Coarse-Grained and Ephemeral Refactorings}
\label{sec: coarse-grained_and_fine-grained_refactorings}

\subsection{Coarse-Grained Refactoring}

\begin{dfn}[Coarse-grained refactoring]
  A refactoring is referred as a CGR if its code changes are recorded across multiple commits rather than a single commit.
\end{dfn}

\begin{figure*}[tb]\centering
  \includegraphics[width=0.7\linewidth,bb=0 0 680 425]{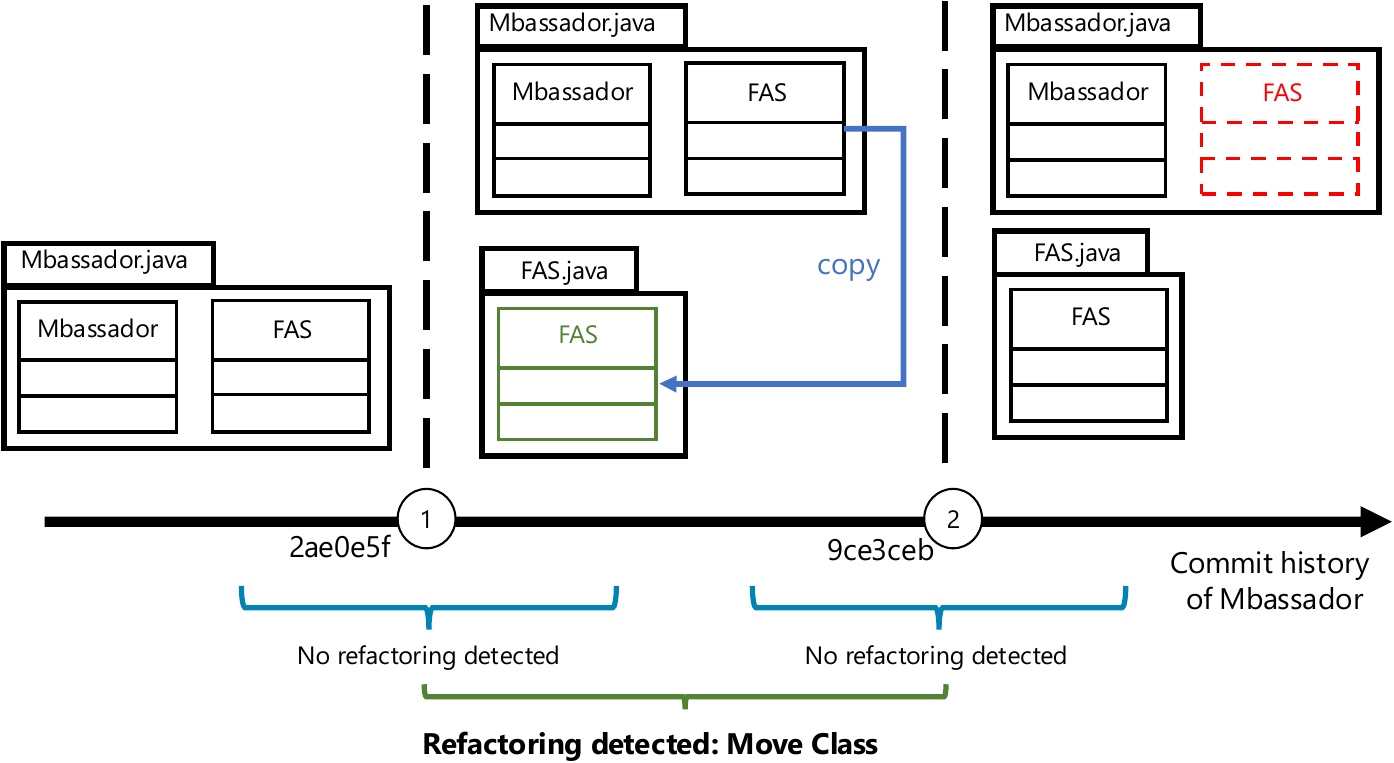}
  \caption{Example of a CGR found in \repository{mbassador}.}
  \label{f:cgr_example}
\end{figure*}

An example of CGR found in open source repository \repository{mbassador}\footnote{\url{https://github.com/bennidi/mbassador/commit/9ce3ceb}} is shown in \cref{f:cgr_example}.
This figure depicts a sample history consisting of two commits, where commit \CommitId{2ae0e5f} is the parent of commit \CommitId{9ce3ceb}.
The intention of the developer, as expressed by these two commits, is to decompose the source file \File{Mbassador.java}, which contains multiple top-level classes, into multiple source files to ensure that each file contains only one top-level class, i.e., applying \MoveClass refactoring.
In the firstly proposed commit \CommitId{2ae0e5f}, the developer copied the implementation of class \Class{FilteredAsynchronousSubscription} (\Class{FAS} hereinafter) in file \File{Mbassador.java} to a newly created file \CodeAdded{FAS.java}.
Then, the developer removed the implementation of that class from the source file \CodeRemoved{Mbassador.java} in the secondly proposed commit.
Overall, she/he moved a class from \Class{Mbassador.java} to a new source file.

The detection based on either of the single commits shown in \cref{f:cgr_example} cannot reveal this kind of refactoring because each commit contains only part of the code changes for detecting \MoveClass refactoring.
However, if we consider the whole changes conducted across these two commits, the refactoring can be correctly detected.

\subsection{Ephemeral Refactoring}

\begin{dfn}[Ephemeral refactoring]
  A refactoring is referred to as an EPR if its code changes are recorded in a single commit, but the code changes in its adjacent commits will break the detection of that refactoring.
\end{dfn}

\begin{figure*}[tb]\centering
  \includegraphics[width=0.7\linewidth]{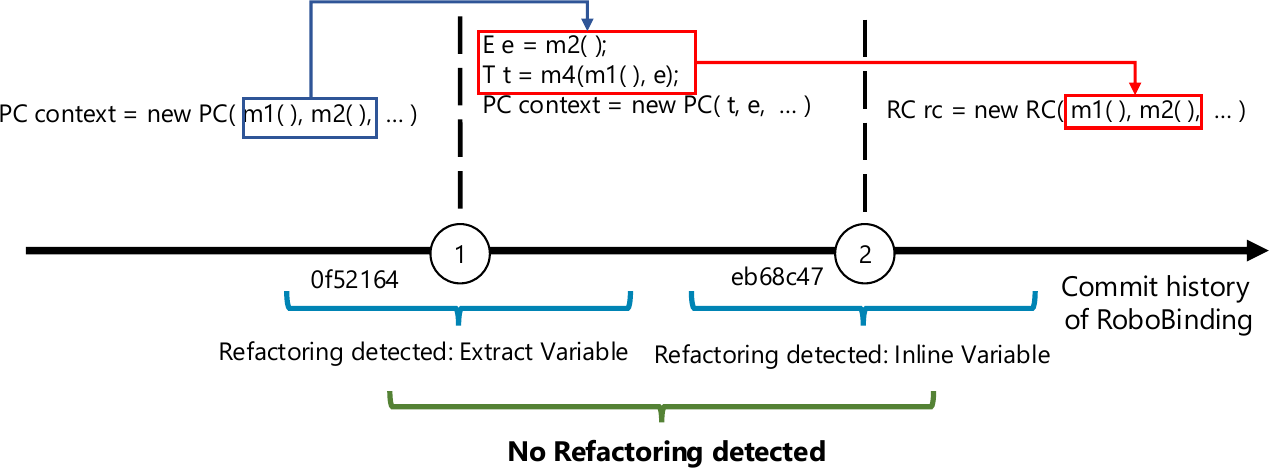}
  \caption{Example of an EPR found in \repository{RoboBinding}.}
  \label{f:epr_example}
\end{figure*}

An example of EPR is found and shown in \cref{f:epr_example}.
This figure shows two commits extracted from the commit history of an open source repository \repository{RoboBinding}\footnote{\url{https://github.com/RoboBinding/RoboBinding/commit/eb68c47}}, where commit \CommitId{0f52164} is the parent commit of commit \CommitId{eb68c47}.
Before the first proposed commit \CommitId{0f52164}, an instance of the \Class{ProcessingContext} (\Class{PC} hereinafter)  class is created using its constructor, which takes three method invocation expressions as arguments.
In the first proposed commit \CommitId{0f52164}, the first two method calls passed to the constructor are extracted into separate variables: \CodeAdded{elements} (\CodeAdded{e} hereinafter)  and \CodeAdded{types} (\CodeAdded{t} hereinafter).
These variables are then used as arguments when instantiating the \Class{PC} object.
This commit introduces two instances of the \ExtractVariable refactoring.
Note that one is not a pure refactoring because a wrapper \Method{m4()} is introduced together with the extraction of the variable \Parameter{t}.
In the subsequent commit \CommitId{eb68c47}, the object type is changed from \Class{PC} to \Class{RoundContext} (hereinafter \Class{RC}). 
Additionally, the previously extracted variables (\CodeRemoved{e} and \CodeRemoved{t}) are inlined back into the constructor arguments.
In this commit, two \InlineVariable refactorings are detected.

The code changes over these two commits are replacing the class \CodeRemoved{PC} with the class \CodeAdded{RC}.
Because the detection of \ExtractVariable refactorings relies on the condition that a new variable is defined, and there is a similarity between its initialization part and some of the removed expressions.
However, the above condition does not hold when we look at the bottom part of \cref{f:epr_example}, and as a result, \ExtractVariable refactorings are not detected at the coarse level.
Similarly, the \InlineVariable refactorings are also not detected at the coarse level.
Since the refactorings \ExtractVariable and \InlineVariable in this example were detected in each commit but were unable to be detected if we look at the whole change of the two commits, they are regarded as EPRs.
The detection on each of the two commits can find the two refactorings.
However, the fact that the overall code change does not contain those refactorings can only be revealed after we inspect the code changes through these two commits.

Note that although the term \emph{coarse-grained refactoring} has been used in other contexts, its meaning differs from ours.
For example, independently of our work, Bibiano et al.~\cite{bibiano2024enhancing} used the same term to describe refactorings that target code elements larger than a method.
Similarly, Mongiovi et al.~\cite{mongiovi2014making} introduced the notion of \emph{small-grained transformation}, referring to code transformation operations for low-level code elements such as ``add method'' or ``remove field'', which are not necessarily refactorings and are not analyzed in relation to commit-based composition.
In contrast, the CGRs and EPRs introduced in this paper refer to newly defined concepts that focus on how refactorings are recorded across commits in a version control system.
The notion of CGR was first proposed in our earlier work~\cite{chen2022impact}, and we continue to use that definition here, extending it by introducing EPRs within the same context.

\section{Related Work}\label{sec:related work}

\subsection{Refactoring Detector}
Detecting refactoring instances performed in software projects can help practitioners be aware of whether and how to track software that requires maintenance and help researchers collect information for empirical studies about refactorings and their effects to build support techniques~\cite{choi2018survey}.
This paper employs a state-of-the-art refactoring detector to mine refactorings from the commit history of software projects.

The input of refactoring detectors comprises two snapshots of the source code.
Based on the main focus of the research targets, the proposed studies can be divided into two types: \textit{release level} and \textit{commit level}.

\subsubsection{Release Level}
Studies of the release level category concentrate on detecting refactorings by utilizing two release versions as the designated snapshots.
Although tools developed in these studies can compare two snapshots of source code to identify refactorings applied between them, they are mainly applied between release versions in their studies.

The \textit{RefactoringCrawler}~\cite{dig2006automated} identifies similar pairs of entities and leverages references among these entities to determine refactoring entities to determine refactoring instances.
This is grounded in the observation that most refactorings include operations of repartitioning source files, which results in source text between different versions of a component being similar.
The Shingles encoding~\cite{broder1997resemblance} from the field of Information Retrieval is adopted to find similar fragments in source files as refactoring candidates.
Then, the refactoring instances can be detected by analyzing the references among the source code entities in each of the refactoring candidates.

The \textit{Ref-Finder}~\cite{kim2010ref,prete2010template} detects refactoring instances of 63 refactoring patterns based on predefined rules.
The detection process involves extracting code elements, their structural dependencies, and the contents of these elements. Subsequently, differences in elements between the two versions are calculated. Finally, refactoring instances are determined based on these calculated differences and predefined rules.

\subsubsection{Commit Level}
This category of refactoring detectors specializes in identifying refactorings within code commits.

Silva et al.\ proposed \textit{RefDiff}~\cite{silva2017refdiff}, which detects refactorings through two phases: source code analysis and relationship analysis.
In the first phase, the tool parses and analyzes the source code to build a model consisting of high-level source code entity information such as types, methods, and fields.
The second phase is to find relationships between elements in the models before and after the code change.
By matching the relationships with predefined rules, refactoring entities can be detected.
Later, \textit{RefDiff} is extended to \textit{RefDiff 2.0}~\cite{silva2020refdiff} to support detecting refactorings in multiple programming languages instead of only in Java.
This version utilizes the \textit{Code Structure Tree}~(CST), a language-agnostic representation of the source code.
The CSTs are matched to analyze the relationship to identify refactoring entities.

Tsantalis et al.\ introduced \textit{RefactoringMiner}~\cite{tsantalis2018accurate}, marking the first refactoring mining tool that operates without the need for any code similarity thresholds.
This tool detects refactorings through Abstract Syntax Tree~(AST)-based statements matching with predefined rules.
The code entities in the two revisions are matched in top-down order based on text similarity.
After that, a set of predefined rules are used to identify refactorings.
The \textit{RefactoringMiner 2.0}~\cite{Tsantalis:TSE:2020:RefactoringMiner2.0} was proposed later, which extended the previous work by supporting the detection of low-level in-method refactoring types.
In addition, they created a refactoring oracle, which contains 7,226 true instances found in 536 commits from 185 open-source projects.

While current refactoring detection studies concentrate on either the release level or the commit level, there is a gap at the intermediate level; refactorings applied across multiple commits remain undetected.

\subsection{Tangled Changes and Composite Refactorings}

In version control systems, a tangled change refers to the practice of committing unrelated or loosely related code changes in a single commit, a phenomenon that occurs frequently during development~\cite{herzig2013impact}.
Tangled changes can hinder the detection of refactorings due to a mismatch between the recorded granularity at which detection is performed and the semantic granularity at which developers actually make changes.
Our study further investigates this mismatch, with a particular focus on changes that span multiple commits.

The tangled change is difficult to understand and revert.
To help developers untangle the tangled changes, Matsuda et al.~\cite{matsuda2015hierarchical} presented a method that restructures changes by tracking source code operations, which encompass refactoring actions, and organizes them hierarchically according to their types provided by an integrated development environment. 
This hierarchy empowers developers to effortlessly adjust the level of change granularity and obtain the resultant changes based on their chosen granularity configuration.
Sothornprapakorn et al.~\cite{sothornprapakorn2018visualizing} propose a visualization approach to support untangling.
The proposed approach organizes tangled changes into a tree structure, leveraging refactoring detection and change relevance calculations to create multiple change groupings within the structure.
The authors conducted an experiment with industrial developers and proved that the tool could help developers understand and decompose the change.

Being a special type of code change, it often occurs in practice that refactorings are mixed with other refactorings or code changes in a single commit~\cite{ge2014towards,ge2017refactoring, alves2014refdistiller}.
Sometimes, such mixed refactorings in a single commit are applied to optimize irrelevant code, while in other instances, they serve as a step towards completing a complicated optimization.
The refactoring detection applied on a single commit granularity, which is the code change recorded granularity, is in discordance with the change granularity, which might be a single refactoring or multiple refactorings distributed over multiple commits.
Our work also investigates the impact of the latter.

G\"{o}rg et al.~\cite{gorg2005detecting} proposed a term of \textit{purity of refactoring}:
a change that contains mixed refactoring and non-refactoring modifications is applied at the same location at the same time as \emph{impure} refactoring.
The mechanism of EPR in our study aligns with this definition.
To detect impure refactorings, a graph search algorithm-based technique is proposed and evaluated on an \textit{Apache} repository, demonstrating its feasibility~\cite{tsutsumi2016graph}.

The refactorings mixed with interrelated refactorings are referred to as \emph{composite refactorings}.
Sousa et al.~\cite{sousa2020characterizing} provide the first formal and unambiguous definition for composite refactorings and provide two heuristics to reveal the characteristics of composite refactorings.
They found that composite refactorings have a considerable effect on code smells: introducing or removing smells, which is contradictory with previous studies~\cite{bavota2015experimental, bibiano2019quantitative, cedrim2017understanding, silva2016we}.
Bibiano et al.~\cite{bibiano2021look} present a composite refactorings catalog, complete with details about side effects, and recommendations on how to decrease them. 
In addition, they enumerate some scenarios where each recommendation can be applied to remove the code smell.
Some composite refactorings are composed of co-occurring refactorings.
Saika et al.~\cite{saika2014kinds} investigate the frequency of co-occurred refactorings by analyzing usage data of the Integrated Development Environment Eclipse.
They conclude that Refactoring pairs~(\Refactoring{Move}, \Refactoring{Rename}), (\Refactoring{Rename}, \Refactoring{Rename}) and (\Refactoring{Extract}, \Refactoring{Move}) appear more frequently than the others.

To further clarify the distinctions among tangled changes, composite refactorings, CGR, and EPR, particularly in terms of granularity and composition, we compare them from two perspectives: granularity and composition.
From the perspective of granularity, tangled changes~\cite{herzig2013impact} and EPR are confined to a single commit, whereas CGRs span across multiple commits.
Composite refactorings are not bound to a specific commit granularity and may occur within a single commit or across multiple commits~\cite{sousa2020characterizing}.
From the perspective of composition, tangled changes consist of changes and unrelated changes, while EPR consists of a single refactoring.
CGR, in contrast, is composed of multiple refactorings or non-behavior-preserved changes as part of refactoring changes distributed across commits.

\section{Detecting Coarse-Grained and Ephemeral Refactorings}
\label{sec: detecting coarse-grained refactorings and ephemeral refactorings}

\subsection{Basic Idea}\label{ss: basic_idea}

\begin{figure}[tb]\centering
  \includegraphics[width=\linewidth]{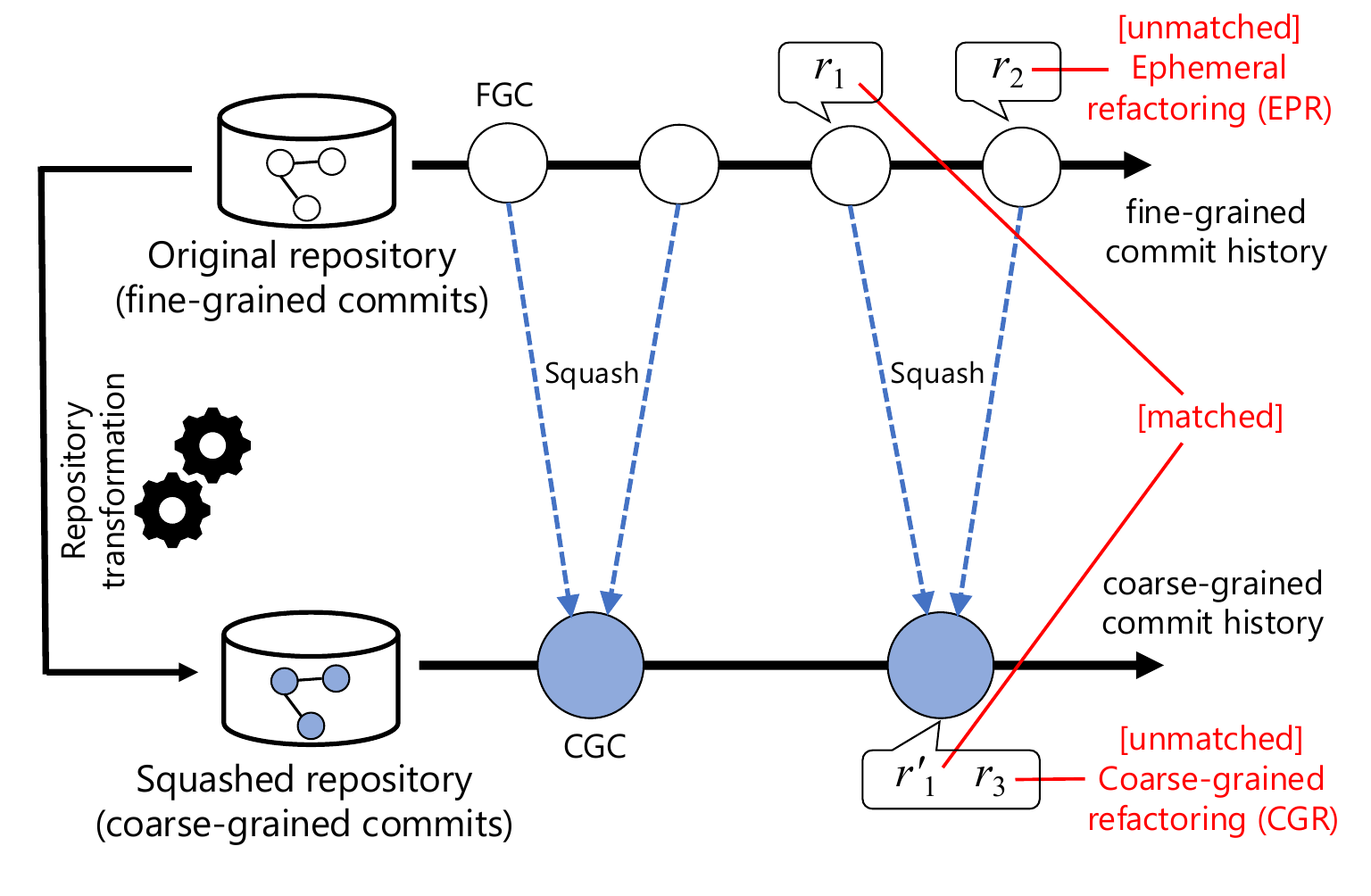}
  \caption{Detection mechanism.}\label{f:matching_mechanism}
\end{figure}

A basic idea to detect CGRs and EPRs is to generate CGCs consisting of the changes in multiple commits in the original repository and to compare the refactoring detection results with the original detection results.
An overview of the mechanism to detect CGRs and EPRs is shown in \cref{f:matching_mechanism}.
We regard the commits stored in the original repository as \emph{fine-grained} in our context.

The FGCs in the original repository are squashed into CGCs through repository transformation.
For each CGC, the fine-grained ones squashed into it are collectively referred to as a \emph{squash unit}.
We tried multiple different strategies to generate the squashed repository so that various kinds of CGCs are examined.
The details are explained in \cref{sec:methodology}.
Then, an existing state-of-the-art refactoring detector is applied to both FGCs and CGCs to detect refactorings.
Refactorings detected from FGCs are referred to as \emph{original refactorings} since those refactorings are detected from the original commit history.
Note that EPRs are a subset of the original refactorings.
While original refactorings include all refactorings detectable at the original commit granularity, EPRs are those that become undetectable once commits are squashed into CGCs.

In the figure, refactorings $r_1$ and $r_2$ are detected from the two commits in the original repository, whereas $r_1'$ and $r_3$ are detected in the commit squashed from those two commits.
By matching the original refactorings and refactorings detected from CGCs, we can classify the detected refactorings into the following three types.
\begin{itemize}
  \item Refactorings that are not detectable from FGCs but from CGCs are identified as CGRs.
  For example, since $r_3$ is detected from the CGC but not from the fine-grained ones, it is regarded as a CGR.
  \item Refactorings that are not detectable from CGCs but from FGCs are identified as EPRs.
  For example, since $r_2$ is only detected from the FGCs but not from the CGC, it is regarded as an EPR.
  \item The remaining refactorings.
  Because $r_1$ and $r_1'$ are matched as the same refactoring, they are considered neither CGR nor EPR.
\end{itemize}

\subsection{Matching Scheme}\label{ss: matching_scheme}

To identify CGRs and EPRs, we compare and match the refactorings detected from the original commit history and those detected from the granularity-changed commit history.
To accurately match two refactorings, it is essential to establish a distinctive \textit{signature} for each refactoring.
This means that two refactoring instances obtained from different commits are considered identical if their signature is identical.

We use RefactoringMiner~\cite{Tsantalis:TSE:2020:RefactoringMiner2.0}, which is the state-of-the-art and widely used refactoring detector, as an example to show its detection result of one refactoring:
\begin{itemize}
  \item 1) refactoring type,
  \item 2) description of how this refactoring is conducted, and
  \item 3) location information of the refactoring target elements, which includes the file paths and line numbers of the refactoring target in both the pre-refactoring and post-refactoring states as well as its role on the refactoring.
\end{itemize}
The 1) \emph{refactoring type} describes the types of refactorings, such as \MoveMethod or \InlineClass.
The 2) \emph{description} textually explains the conducted refactoring, consisting of the type of the refactoring being applied and its details such as the type and name of the refactored code elements or the file path of the class containing the refactored code elements.
Unfortunately, comparison by types or descriptions is insufficient to distinguish small refactorings, 
in cases where two refactorings of the same type are conducted together or code elements with the same type and same name but are located in different methods in the same file are refactored together.
The richest information source is the 3) \emph{location information}, which can help to solve the above issue.
By using both the file path and the line number, one code element can be uniquely located.
In this approach, the refactorings are characterized by the role and location of the refactored code elements.

\begin{figure}[tb]\centering
  \includegraphics[width=\linewidth]{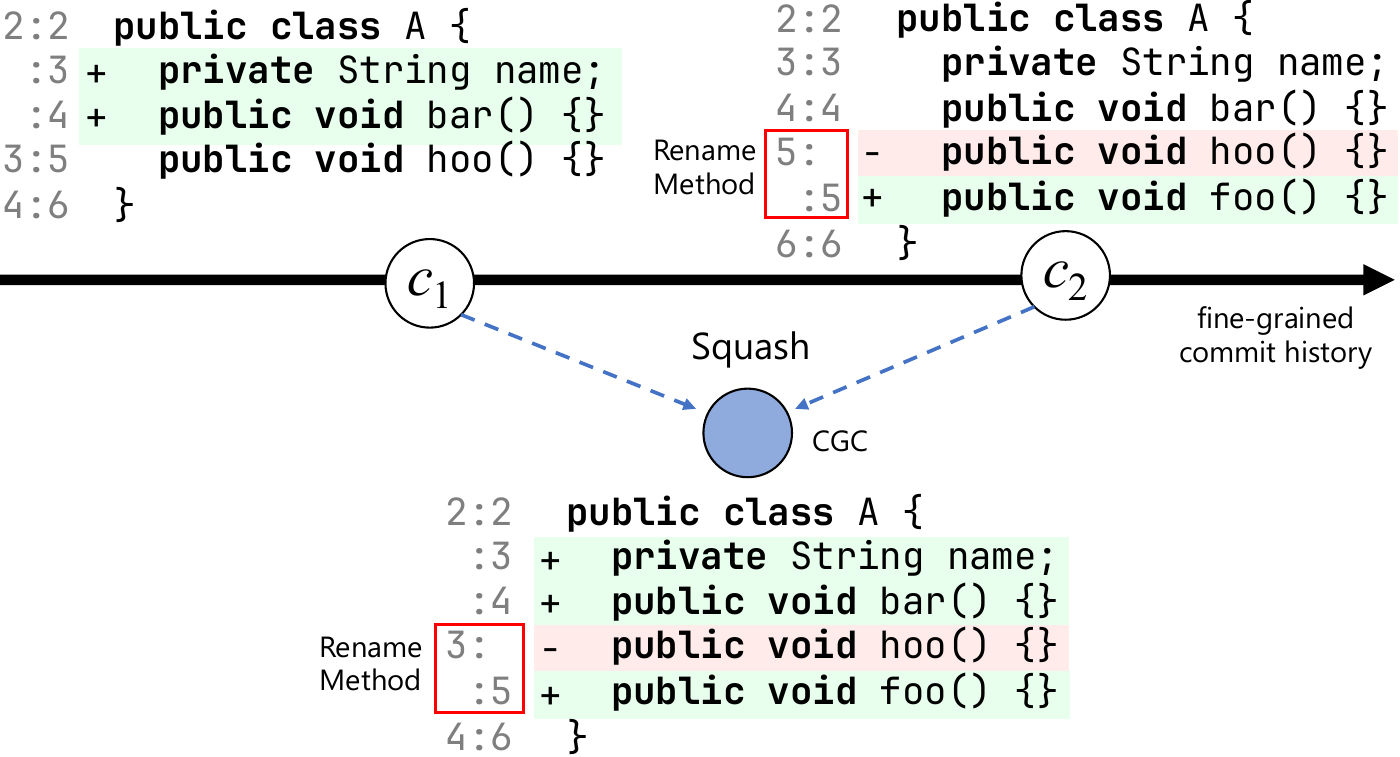}
  \caption{Example of location change because of granularity change.}
  \label{f:location_change}
\end{figure}
However, changing granularity may impact the location (line number) where refactoring is performed, which in turn affects the matching process.
An example is shown in \cref{f:location_change}.
The code change in the FGC $c_2$ is to rename the method from \CodeRemoved{hoo} to \CodeAdded{foo}.
It is detected as a \RenameMethod, and it was applied to the code element in Line~5.
In the previous commit $c_1$, two lines of code are added above the method declaration of \Method{hoo()}.
In the squashed CGC of the above two commits, a refactoring \RenameMethod is detected.
It is equal to the \RenameMethod detected from FGCs, and they should be matched as the same refactoring.
However, their locations differ: the one in the squashed commit is detected to be applied on Line~3 instead of Line~5 because of the code inserted in the commit $c_1$.
Therefore, checking the exact equivalence of the location where the refactorings are applied is insufficient to match two refactorings.

We trace the refactoring location in order to match refactorings.
We name the refactoring targeted source code snippets at the version before applying the refactoring as the \emph{refactoring targets}.
For example, the refactoring target for \RenameMethod is the line of the code containing the declaration of the method undergoing the renaming process.
The trace of refactoring is to find out the commit that the refactoring target is last modified and the line number of the first line of that code in the found commit.
The first line here refers to the line with the smallest line number in the location information of the refactoring detection result for the refactoring target.
In case of refactoring detection result involves multiple code elements,
We choose the first line of the first code element as it is usually the primary element in the detection result.
We use the first line of refactoring target to trace because it is the class name, method header, or code statement, which can represent the code block of being refactored.

Following the techniques employed in violation tracing studies~\cite{avgustinov2015tracking,hanam2014finding}, we use the \emph{line origin}, the line of the commit when the target line was introduced, for the key factor of the signature.
The first line of the refactoring target is typically the signature of the refactored code element, which effectively represents the element.
We apply \textit{git-blame} on the first line of the refactoring target to specify the origin commit and the line at the origin commit.
Although tracing techniques such as CodeTracker~\cite{jodavi2022accurate, hasan2024refactoring}, which focus on Java code and are refactoring-aware with high tracing accuracy, are available, we chose not to adopt them.
This decision was based on two considerations.
First, extending the tools to support all types of refactorings investigated in this study would require additional effort, which is beyond the scope of our work.
Second, we wanted to design the detection framework to be as language-agnostic as possible so that interested followers could easily replicate it in other programming languages, whereas techniques such as CodeTracker are specific to Java.
As a result, we employed \textit{git-blame}, a more general technique that can be applied across a wide range of programming languages.
Two refactorings are compared to see if they reach the same line at the same origin commit.
The details are introduced in \cref{subsec:refactoring_location_tracing}.

Note that before the comparison and matching, we have a preprocessing step to remove the comments from the source code that may interfere with identification accuracy.
Some refactoring detection results regard the first line of a class or a method as the Javadocs or comments above the code, instead of the line declaring the class or method.
Therefore, the changes to comments, such as the addition or removal of Javadoc comments, may influence the refactoring signature.
The situation where the same refactoring is no longer matched due to modifications irrelevant to the behavior of the source code is not desired.
Therefore, we excluded such comments in advance as a pre-processing step to eliminate their influence.
The details are introduced in \cref{subsec:preprocesing}.

Compared with our previous work~\cite{chen2022impact}, we enhanced the matching mechanism by designing a more delicate refactoring \textit{signature}.
In our previous work, only the refactoring type is used for matching, which may cause false negatives where some CGRs are omitted because of having the same type with refactorings in the original history.
By using both the traced location and refactoring type, we are able to uncover more CGRs and EPRs with higher accuracy.

\section{Methodology}\label{sec:methodology}

\subsection{Overview}\label{subsec:overview}

\begin{figure*}[tb]\centering
  \includegraphics[width=0.95\linewidth]{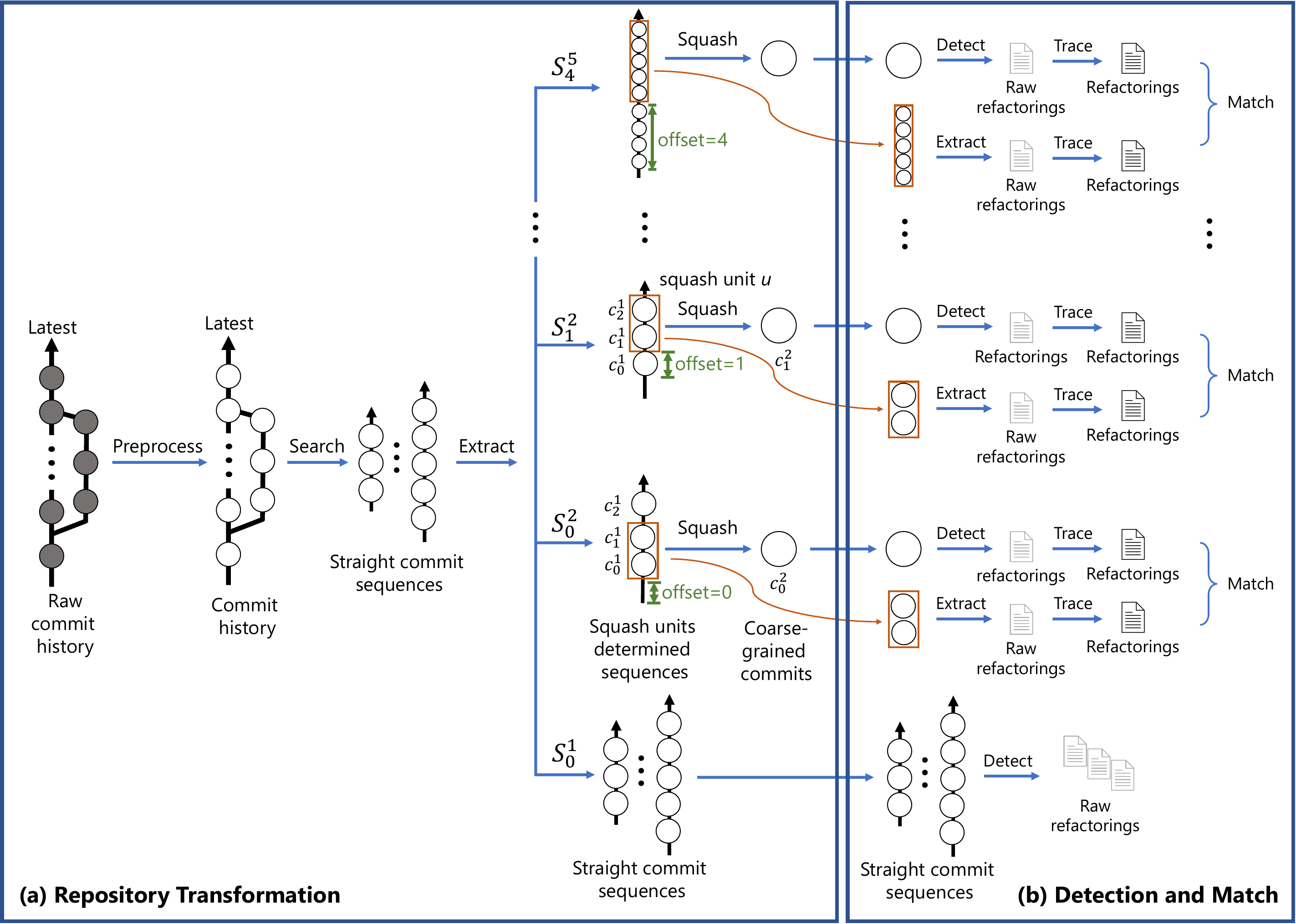}
  \caption{Study overview.}
  \label{f:study_overview}
\end{figure*}
The overview of our study procedure is shown in \cref{f:study_overview}.
Our procedure can be divided into two phases: (a) \emph{Repository Transformation} and (b) \emph{Detection and Match}.

In the repository transformation phase, the input is the raw Git-based commit history extracted from a repository.
Through preprocessing, which removes the Javadocs and comments, we can obtain a commit history that excludes changes related with comments.
Searched from commit history, we can obtain \emph{straight commit sequences}, each containing consecutive commits on the same branch.
The \emph{squash units} can be determined on the straight commit sequences, and each of them is squashed into a CGC.

In the detection and match phase, for each pair of CGC and the FGCs that are squashed into that CGC, refactorings are detected and matched.

\subsection{Preprocessing}\label{subsec:preprocesing}

\begin{figure}[tb]\centering
  {\footnotesize
  \includegraphics[width=\linewidth]{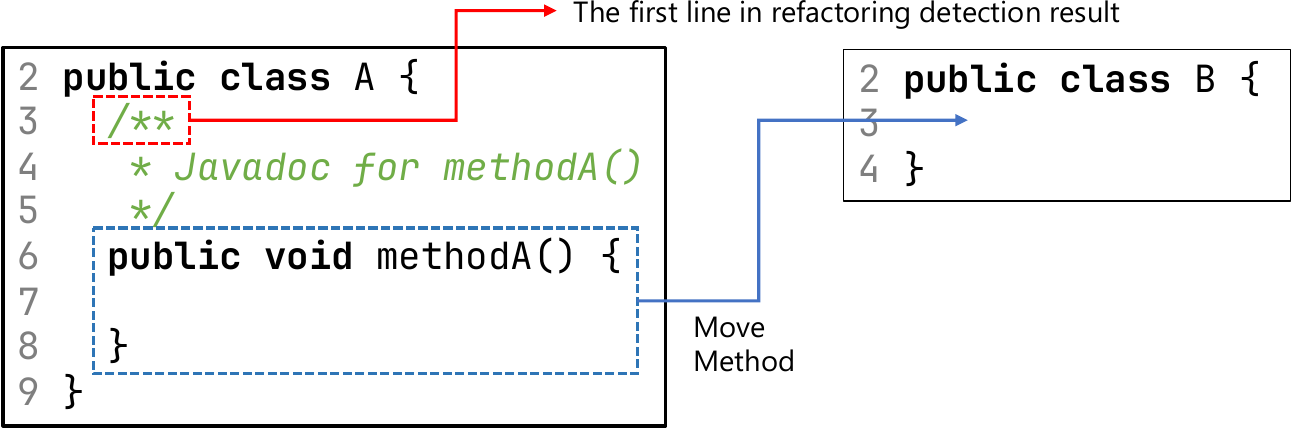}\\
  (a) Before preprocessing.\\\vspace{2em}
  \includegraphics[width=\linewidth]{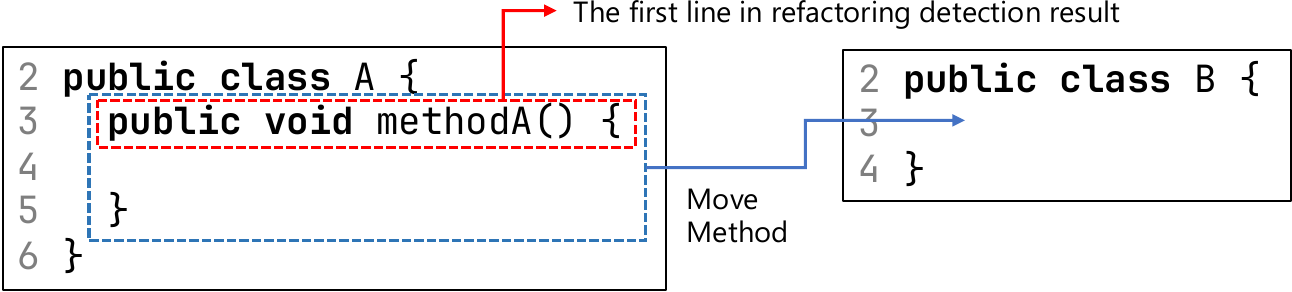}\\
  (b) After preprocessing.}
  \caption{Example of preprocessing.}\label{f:preprocessing_example}
\end{figure}
As introduced in \cref{ss: matching_scheme}, some refactoring detectors regard the first line of a class method as the Javadocs or comments above the code.
We remove those comments in the preprocessing phase to prevent the code changes in comments from affecting the tracing results.
An example showing the above procedures is depicted in \cref{f:preprocessing_example}.
The refactoring is to move the method \Method{methodA()} from class \Class{A} to class \Class{B}.
Before the preprocessing, the first line of the refactoring target is regarded as beginning at Line~3, which is the Javadoc comment for method \Method{methodA()}.
After the preprocessing, the comment is removed, and the first line of the refactored code snippet specifies the method signature of the method being moved.

\subsection{Extracting Straight Commit Sequences}\label{subsec:search_straight_commit_sequences}

From the preprocessed Git-based commit history $H$, which is a set of FGCs $C$ ($\subseteq \CommitSet$), where $\CommitSet$ is the universal set of commits, straight commit sequences $h \subseteq C$ can be extracted.
Each straight commit sequence $h$ consists of consecutive FGCs on the same branch that exclude \textit{merge commits}, which have more than one parent, and \textit{branch sources}, which have more than one child.
Merge commits are excluded to avoid duplicate detection of refactoring in the later phase, and branch sources are excluded for simplicity when extracting squash units.
Here, $\seqs(H)~(\subseteq 2^C)$ denotes the set of all the possible straight commit sequences extracted from $H$.

\subsection{Determining Squash Units}\label{subsec:squash_units_determination}

A \emph{squash unit} $u$ ($\subseteq C$) is a set of multiple adjacent FGCs that are squashed into a single CGC.
Here, if a commit is the parent or child of another commit, these two commits are considered adjacent.
The adjacent commits are shown as circles next to each other in \cref{f:study_overview}.

Different strategies labeled $S_o^\Level$ (for appropriate values of $o$ and $\Level$) are used to extract squash units from straight commit sequences.
Here, the \emph{granularity level} $\Level\ (\geq 1)$ specifies the size of the squash units, and straight commit sequences are divided into multiple squash units of the specified size.
Because each unit is squashed into one CGC, this level also expresses the granularity level of the CGCs to be generated.
The granularity level $\Level=1$ exactly produces original FGCs.
The \emph{offset} $0\leq o\leq$ $\Level-1$ is the number of commits to be skipped from the beginning of the given straight commit sequence when extracting the squash units to adjust which commits will be merged.
For example, the commit $c_1^1$ in \cref{f:study_overview} is squashed together with $c_0^1$ when strategy $S_0^2$ is used, whereas it is squashed together with $c_2^1$ when strategy $S_1^2$ is used.

To be more specific, for the straight commit sequence $h = \{ c_1^1, c_2^1, \dots, c_n^1 \}$, containing $n$ consecutive FGCs, the squash units extracted from $h$ with the given strategy $S_o^\Level~(\Level \geq 2)$ can be presented as:
\begin{align*}    
   \units_{S_o^\Level}(h) = \{ u_{o + k\Level} \mid 0 \leq k \leq (n - \Level - o) / \Level \}
\end{align*}
where $u_x = \{ c_i^1 \mid x \leq i < x + \Level \}$ is the squash unit beginning from the commit $c^1_x$ determined by strategy $S_o^\Level$.
We regard the size of a squash unit as its \emph{granularity}.
We may write the granularity level as the superscript of the unit, e.g., $u^\Level$ ($\Level = |u|$).

\begin{figure}[tb]\centering
  \includegraphics[width=8cm]{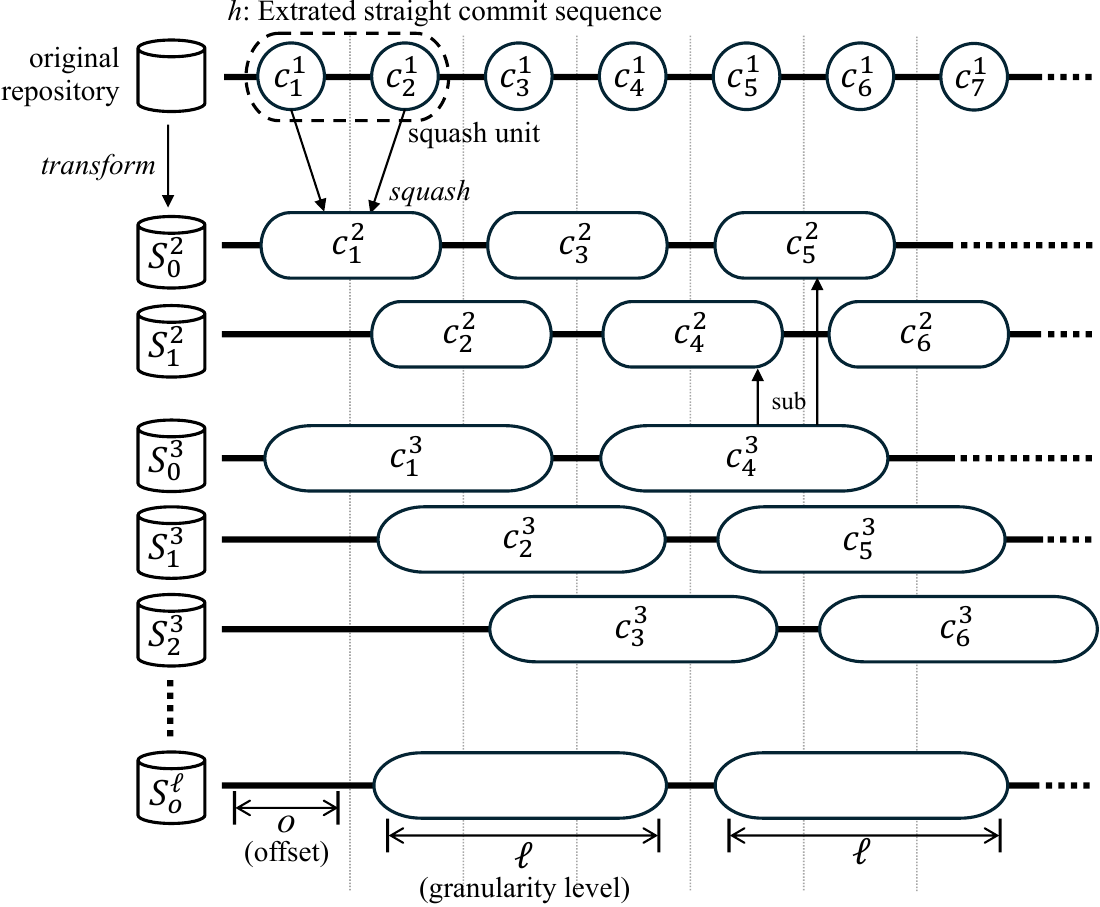}
  \caption{Comparison of different strategies.}\label{f:strategy-comparison}
\end{figure}
The difference of the generated commits according to different strategies is illustrated in \cref{f:strategy-comparison}.
Different squash units are created according to the strategy with a specific granularity level and offset, and different squashed commits are generated from these squash units.
By covering all possible offsets at each granularity level, all possible squash units could be enumerated.
For example, at the granularity level of $\Level=3$, three different offsets $o \in \{0,1,2\}$ are used.

\subsection{Changing Granularity}\label{subsec: granularity_change}

To change the granularity of commit history, we squash multiple commits into a single one to integrate the code changes distributed in multiple revisions.
The tool \textit{git-stein}\footnote{\url{https://github.com/sh5i/git-stein}}~\cite{shiba-jssst202211}, which can squash the FGCs in all the squash units extracted from the repository into CGCs, is used to perform the granularity transformation.
The output of the tool is another repository whose commit history is composed of CGCs.
For each squash unit $u^\Level$ determined in \cref{subsec:squash_units_determination}, $c^\Level = \sq(u^\Level)$ represents squashing all the $\Level$ commits in $u^\Level$ into a single CGC $c^\Level$.
Similarly to squash units, we may specify the granularity level as the superscript of the commit.

A squashed commit of a certain granularity level can cover several commits of finer granularity levels.
In case where a commit $c^\Level$ is generated by squashing a unit $u^\Level = \{c^1_i, \dots, c^1_{i+\Level-1} \}$, its \emph{sub commits} and \emph{super commits} can be defined as follows:
\begin{align*}
   \Sub(c^\Level) &= \{ \sq(u) \mid u \subsetneq u^\Level \} \\
                  &= \{ \sq(\{c^1_j, \dots, c^1_{j+\Level'-1}\}) \mid
                    1 \leq \Level' < \Level \wedge
                    i \leq j < \Level - \Level' \}, \label{eq:sub} \\
   \Sup(c^\Level) &= \{ \sq(u) \mid u \supsetneq u^\Level \} \\
                  &= \{ \sq(\{c^1_j, \dots, c^1_{j+\Level'-1}\}) \mid
                    \Level < \Level' \wedge
                    j \leq i \wedge
                    i+\Level \leq j+\Level' \}.
\end{align*}
For example, $c^2_4$ and $c^2_5$ in \cref{f:strategy-comparison} are sub-commits of $c^3_4$.
Also, the original FGCs ($c^1_4, c^1_5, c^1_6$) are also regarded as sub-commits of $c^3_4$.

All the squashed commits at granularity level $\Level$ extracted from the commit history $H$ can be represented as:
\begin{align*}
    C^\Level(H) = \left\{ \sq(u) \relmiddle|
       u \in \bigcup_{0 \leq o \leq \Level-1}
             \bigcup_{h \in \seqs(H)}
             \units_{S_o^\Level}(h) \right\}.
\end{align*}

Note that the granularity change can be applied at once in a repository scale; applying repository transformation of strategy $S$ to the whole repository, we can obtain another repository consisting of CGCs.
The relationship between a squash unit and its squashed CGC can be determined by associating the original FGCs with CGCs in the transformed repository.

\subsection{Detecting Refactorings}
Refactoring detection is conducted on both the commit history before and after the granularity transformation.
The refactorings detected from a commit $c$ can be defined as:
\begin{align*}
    R = \refc(c).
\end{align*}
Here, $R \subseteq \mathcal{R}$ is a set of refactorings, where $\mathcal{R}$ denotes the universal set of refactorings.
For each refactoring $r \in R$, 
  $\Type{r}$ denotes its type, 
  $\Sig{r}$ denotes the refactoring target signature, and
  $\Location{r}$ denotes the traced location whose detail is explained in \cref{subsec:refactoring_location_tracing}.
We consider the first element in the pre-refactoring version of the code snippet in the detection result of RefactoringMiner, to be the primary element of the refactoring and use its location as the target for tracing, i.e., refactoring targets as introduced in \cref{ss: matching_scheme}.

\subsection{Refactoring Location Tracing and Match}\label{subsec:refactoring_location_tracing}

\begin{figure*}[tb]\centering
  \includegraphics[width=0.6\textwidth]{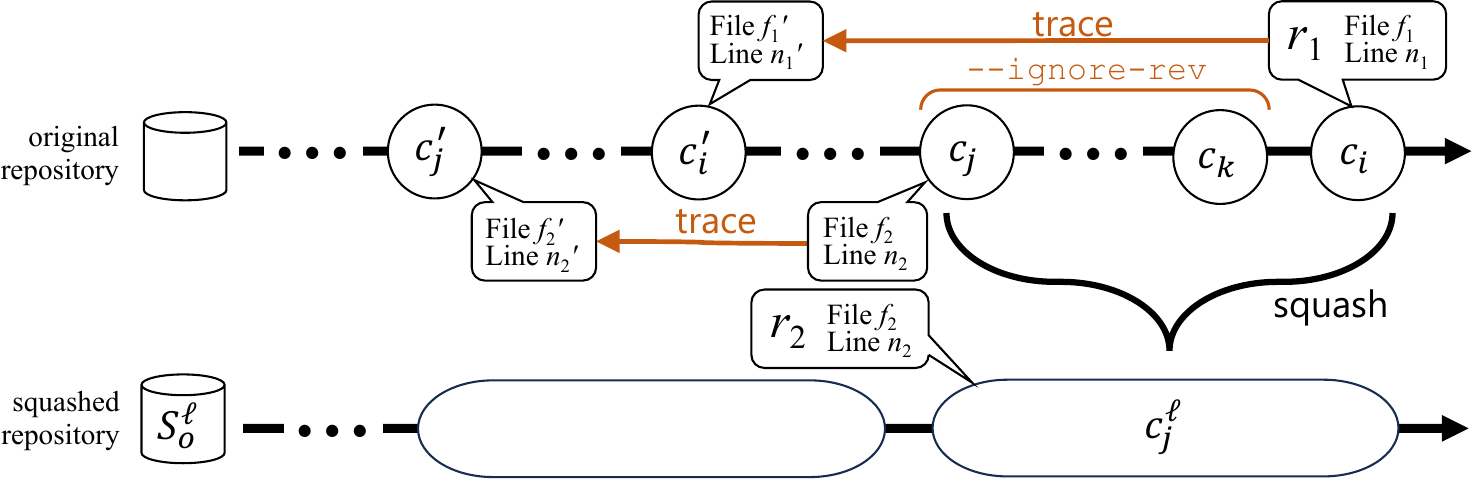}
  \caption{Refactoring location tracing.}\label{f:refactoring_location_trace}
\end{figure*}

As discussed in \cref{ss: matching_scheme}, the match of refactorings detected in the CGCs and those in FGCs requires location tracing to mitigate the change in refactoring location brought by granularity change.

The overview of the tracing procedure is depicted in \cref{f:refactoring_location_trace}.
In the commit history shown, the $\Level$ number of FGCs $c_j, \dots, c_i$ are squashed into a CGC $c_j^\Level$ at the granularity level of $\Level$.
Here, a refactoring $r_1$ at Line $n_1$ on File $f_1$ is detected from the FGC $c_i$ and another refactoring $r_2$ at Line $n_2$ on File $f_2$ is detected from a CGC $c^\Level_j$.
Our aim is to trace the location, where the refactored code snippet is modified in the nearest parent commit of the being squashed commits.

Following the techniques employed in violation tracing studies~\cite{avgustinov2015tracking,hanam2014finding}, we implemented the trace mechanism using \textit{git-blame}.
Applying \textit{git-blame} on a certain line of code can reveal the information about the last modification of that line.
The information includes the commit hash, author, date, location, which is the line number, and the contents.
The target of our tracing is the commit hash, file path, and the line number of the refactored code.
The first line of the refactoring target is typically the signature of the refactored code element, which effectively represents the element.
We apply \textit{git-blame} on the first line of the refactoring target of $r_2$.
Since $r_2$ is detected in the squashed repository, we perform the tracing on the original repository to enable comparison with results from the original commit history.
In this case, we run \textit{git-blame} at  $n_2$ on $f_2$ at  $c_j$, which is the oldest commit in the squash unit of $c^\Level_j$.
In the figure, the result of \textit{git-blame}-based tracing revealed that the origin of the refactoring target is found in $c_j'$ at Line $n_2'$ on File $f_2'$, indicating the most recent modification of the refactored code element at this line.

If the commits being squashed contain code changes applied to the refactoring target, the tracing result will be found in one of these commits. 
However, since these commits are squashed into a CGC, they cannot serve as the tracing result for refactorings in the CGC.
To match the refactorings from both coarse-grained and fine-grained, in tracing $r_1$, we trace the commit, excluding those being squashed, that last modified the first line of the refactoring target.
This is achieved by utilizing \textit{git-blame} with the \texttt{--ignore-rev} option to ignore the code changes in the squashed commits.
In this case, we run \textit{git-blame} at  $n_1$ on $f_1$ at commit $c_i$ while ignoring the commits of $c_j, \dots, c_k$.
In the figure, the trace result is in $c_i'$ at Line $n_1'$ on File $f_1'$.

Two refactorings $r_1$ and $r_2$ are matched as the same refactoring ($r_1 \sim r_2$) if they conform to the same conditions:
  1) having the same refactoring type ($\Type{r_1} = \Type{r_2}$),
  2) having the same refactoring target signature ($\Sig{r_1} = \Sig{r_2}$), and
  3) their refactoring targets are traced to the same location at the same commit ($c_i' = c_j' \wedge f_1' = f_2' \wedge n_1' = n_2'$).

\subsection{Comparison and Matching}\label{comparison_and_matching}
To identify CGRs and EPRs, comparison and matching are performed among refactorings detected from different levels of CGCs and from FGCs.

To obtain CGRs, refactorings detected from CGCs are compared with those detected from CGCs at a lower granularity level and FGCs.
The comparison with refactorings from lower granularity level CGCs is necessary because some CGRs at a lower granularity level may also be detected at a higher granularity level.
To be more specific, a CGR $r$ detected from CGC $c_u = \sq(u)$, which is squashed by a squash unit consisting of FGCs $u = \{c_1, c_2\}$, may also be detected from another CGC $c_{u'} = \sq(u')$, which is squashed by FGCs $u' = \{ c_1, c_2, c_3 \}$, if the code change in $c_3$ does not disrupt the detection of $r$.
The smallest squash unit where the CGR is detected is considered as its granularity level.
The CGRs deduced from a squashed commit $c = \sq(u)$ are defined as follows:
\begin{align*}
  \CGR(c) &= \left\{ r \in \refc(c) \relmiddle|
    \nexists r' \in \bigcup_{c' \in \Sub(c)} \refc(c')
      \bullet r \sim r' \right\}.
\end{align*}

To obtain EPRs, the traced refactorings detected from FGCs are compared with the refactorings detected from CGCs at a higher granularity level.
The granularity level of the squash unit that contains the least FGCs where the EPR is detected in FGC but not in CGC is considered as its granularity level.
The EPRs at granularity $\Level$ deduced from a FGC $c$ are defined as follows:
\begin{align*}
  \EPR^\Level(c) &= \left\{ r \in \refc(c) \relmiddle|
    \nexists r' \in \bigcup_{c' \in \sup(c) \cap C^\Level(H)} \refc(c') \bullet r \sim r' \right\} \\
    & ~~~~~~  \setminus \bigcup_{\Level' < \Level} \EPR^{\Level'}(c).
\end{align*}

Note that the above definition of CGR differs from the one used in our previous study~\cite{chen2022impact}.
In the previous study, we defined a refactoring $r$ detected from a CGC $c=\sq(u)$ squashed from a squash unit $u$ as coarse-grained if and only if no refactoring of its type was found in the detected refactorings from each FGC in $u$:
\begin{align*}
    \CGR'(c) &= \{ r \in \refc(c) \mid 
        \nexists r' \in \bigcup_{c \in u} \refc(c) \bullet
        \Type{r} = \Type{r'\!} \}.
\end{align*}
There are two differences between the two definitions.
The first difference is that, in this study, we improved the accuracy using a more sophisticated signature to identify the same refactorings among different granularity levels, instead of using only the type of refactorings as the signature. 
The second is that, in this study, refactorings found at the granularity level $\Level$ were not regarded as CGRs if they were also found at any finer granularity than $\Level$, which improves the conceptual validity.

Our implementation is publicly available online\footnote{\url{https://github.com/MashiroCl/GranulRef}}.

\section{Empirical Study}\label{sec: empirical_study}

\subsection{Research Questions}
We set five research questions~(RQs) to investigate the impact of the granularity change in refactoring detection from the aspects of 1) the appearance frequency of CGRs and EPRs (\RQ{1} and \RQ{5}), 2) type analysis of CGRs and EPRs (\RQ{2} and \RQ{5}), 3) the cause of CGRs and EPRs (\RQ{3} and \RQ{5}), and 4) the relationship between commit messages and CGR (\RQ{4}).
The five RQs are as follows:

\def\RQone{How frequently do CGRs appear because of granularity change?}
\def\RQtwo{What are the types of CGRs?}
\def\RQthree{How are CGRs introduced in development?}
\def\RQfour{Do commit messages suggest the existence of CGRs?}
\def\RQfive{What are the features of EPRs?}

\begin{description}
    \item[\textbf{\RQ{1}:}] \RQone
    \item[\textbf{\RQ{2}:}] \RQtwo
    \item[\textbf{\RQ{3}:}] \RQthree
    \item[\textbf{\RQ{4}:}] \RQfour
    \item[\textbf{\RQ{5}:}] \RQfive
\end{description}
The details of each RQ are introduced below.

\subsection{Data Collection}\label{subsec: dataset_collection}
\begin{table*}[tb]\centering
\caption{Dataset used}\label{t:dataset}
{\scriptsize\begin{tabular}{lrrrrrrrrl} \hline
 & & \# commit & \# involved & Avg. seq. & \multicolumn{4}{c}{Average \# squash units} & \\
 Repository & \!\!\!\!\!\# commits & sequences & commits & length & $\Level=2$ & $\Level=3$ & $\Level=4$ & $\Level=5$ & Description/domain \\
 \hline
\repository{mbassador} & 342 & 84 & 295 & 3.51 & 99.00 & 58.33 & 39.75 & 29.60 &  Event bus\\
\repository{retrolambda} & 530 & 54 & 506 & 9.37 & 224.50 & 140.00 & 99.50 & 75.40 & Backport of lambda expression\\
\repository{seyren} & 640 & 250 & 498 & 1.99 & 104.50 & 49.33 & 29.75 & 18.20 & Dashboard\\
\repository{android-async-http} & 899 & 265 & 754 & 2.85 & 226.00 & 128.67 & 82.00 & 58.20 & HTTP client\\
\repository{javapoet} & 937 & 457 & 621 & 1.36 & 74.50 & 23.00 & 9.75 & 6.40 & Source file generator \\
\repository{sshj} & 1,045 & 155 & 976 & 6.30 & 396.00 & 243.00 & 169.00 & 127.20 & SSH library\\
\repository{RoboBinding} & 1,088 & 301 & 956 & 3.18 & 306.00 & 164.33 & 97.50 & 64.80 & Data binding framework\\
\repository{giraph} & 1,138 & 17 & 1,132 & 66.59 & 556.00 & 367.33 & 274.00 & 218.00 & Graph processing system\\
\repository{jeromq} & 1,470 & 579 & 1,063 & 1.84 & 200.00 & 99.00 & 58.75 & 40.00 & Messaging library\\
\repository{jfinal} & 1,655 & 144 & 1,595 & 11.08 & 717.50 & 459.67 & 332.50 & 256.60 & Web framework\\
\repository{zuul} & 1,689 & 357 & 1,517 & 4.25 & 551.00 & 326.67 & 224.00 & 164.40 & Gateway service\\
\repository{baasbox} & 1,706 & 662 & 1,191 & 1.80 & 226.00 & 105.67 & 54.00 & 35.20 & Backend server\\
\repository{spring-data-rest} & 1,755 & 34 & 1,740 & 51.18 & 851.00 & 563.00 & 419.50 & 333.20 & RESTful data access\\
\repository{truth} & 2,006 & 205 & 1,868 & 9.11 & 809.00 & 514.00 & 372.25 & 290.80 & Java assertions\\
\repository{rest-assured} & 2,309 & 142 & 2,243 & 15.80 & 1,042.50 & 671.33 & 492.25 & 385.80 & REST service testing\\
\repository{helios} & 2,458 & 1,166 & 1,727 & 1.48 & 218.00 & 99.33 & 53.75 & 33.80 & Container orchestration framework\\
\repository{cascading} & 2,528 & 190 & 2,434 & 12.81 & 1,113.50 & 694.00 & 489.25 & 368.00 & Data processing\\
\repository{goclipse} & 2,925 & 651 & 2,496 & 3.83 & 865.50 & 502.00 & 336.50 & 245.40 & IDE for Go language\\
\repository{HikariCP} & 2,929 & 389 & 2,746 & 7.06 & 1,154.00 & 711.67 & 492.25 & 369.00 & JDBC connection\\
\repository{rest.li} & 2,931 & 67 & 2,908 & 43.40 & 1,415.00 & 930.33 & 690.75 & 547.20 & REST framework\\
\repository{hydra} & 2,958 & 487 & 2,749 & 5.64 & 1,098.50 & 633.33 & 420.00 & 299.40 & Distributed data processing\\
\repository{blueflood} & 3,152 & 910 & 2,561 & 2.81 & 756.50 & 402.00 & 249.00 & 168.60 & Data processing\\
\repository{PocketHub} & 3,512 & 525 & 3,211 & 6.12 & 1,314.00 & 840.67 & 613.00 & 480.40 & Android app\\
\repository{xabber-android} & 4,264 & 376 & 4,095 & 10.89 & 1,844.50 & 1,170.00 & 840.25 & 647.80 & XMPP client for Android\\
\repository{morphia} & 4,499 & 731 & 4,152 & 5.68 & 1,694.50 & 1,087.33 & 790.75 & 614.80 & Java MongoDB ORM\\
\repository{redisson} & 10,014 & 2,907 & 8,638 & 2.97 & 2,604.50 & 1,375.33 & 842.75 & 559.40  & Redis client\\
\repository{Activiti} & 11,135 & 3,017 & 9,543 & 3.16 & 3,056.00 & 1,781.00 & 1,213.75 & 902.60 & Business Process Management Platform\\
\repository{processing} & 13,211 & 2,106 & 12,299 & 5.84 & 4,944.50 & 3,076.33 & 2,192.75 & 1,682.60 & Code learning platform\\
\repository{checkstyle} & 14,360 & 40 & 14,342 & 358.55 & 7,148.00 & 4,758.33 & 3,565.50 & 2,849.00 & Code quality linter\\
\repository{libgdx} & 15,433 & 4,443 & 13,177 & 2.97 & 4,021.00 & 2,302.00 & 1,562.50 & 1,159.00 & Game development framework\\
\repository{cgeo} & 18,941 & 5,336 & 16,103 & 3.02 & 4,938.50 & 2,666.67 & 1,666.75 & 1,145.80 & Client for geocaching\\
\repository{JGroups} & 20,367 & 1,191 & 19,782 & 16.61 & 9,242.50 & 6,035.33 & 4,453.75 & 3,516.40 & Messaging library\\
\hline
Total & 154,826 & 28,238 & 139,918 & 4.95 & \\
\hline
\end{tabular}}
\end{table*}

The repositories that we selected are from a dataset collected by Silva et al.~\cite{silva2016we}, containing 124 GitHub-hosted Java projects.
These repositories contain refactorings, and some of them have been identified by RefactoringMiner, studied, and confirmed by researchers.
Given the variety of coding conventions and practices in different domains of repositories, the refactorings applied in each domain may be different.
To mitigate the above bias, we chose a total of 32 repositories from the 124 repositories with consideration of the variety in the projects' domains.
The 19 repositories used in our previous study~\cite{chen2022impact} were included, and the remaining 13 repositories were randomly selected, each with distinct domains.
The detail of our dataset is shown in \cref{t:dataset}. 
Each column represents the repository name, the number of commits to be processed in the experiment, the number of straight commit sequences, the number of commits involved in the straight commit sequences, the average length of straight commit sequences, the average number of squash units at each granularity level from 2 to 5, the number of refactorings detected, and the domain of the repository.
Note that in \cref{subsec:squash_units_determination}, we introduced that different strategies determined by different offset values are used for determining squash units at a certain granularity level, e.g., there are two strategies at $\Level=2$, three strategies at $\Level=3$.
The average number of squash units is calculated using the total number of squash units under different strategies to divide the number of strategies at that granularity level.
The domain of repositories encompasses web frameworks, Android apps, performance toolkits, plugins, and more.
The state-of-the-art refactoring detector \textit{RefactoringMiner 3.0.4}~\cite{Tsantalis:TSE:2020:RefactoringMiner2.0} was used to detect refactorings.
As described in \cref{subsec:search_straight_commit_sequences}, merge commits and branch commits were excluded.
The number of commits after this exclusion ranged from 295 to 19,782.
The average lengths of straight commit sequences range from 1.36 to 358.55.

\begin{table*}[tb]\centering
\caption{Detected CGRs and EPRs}\label{t:dataset_granularity}
{\scriptsize\begin{tabular}{lr|rrrr|rrrr}\hline
 & \# Original & \multicolumn{4}{c|}{\# CGRs}  & \multicolumn{4}{c}{\# EPRs} \\
Repository & refactorings & $\Level=2$ & $\Level=3$ & $\Level=4$ & $\Level=5$ & $\Level=2$ & $\Level=3$ & $\Level=4$ & $\Level=5$\\
\hline
\repository{mbassador} & 1,019 & 19 & 22 & 11 & 2 & 14 & 2 & 17 & 6\\
\repository{retrolambda} & 855 & 64 & 26 & 39 & 13 & 114 & 61 & 27 & 10\\
\repository{seyren} & 427 & 3 & 1 & 0 & 0 & 26 & 13 & 2 & 1\\
\repository{android-async-http} & 1,092 & 26 & 50 & 0 & 1 & 45 & 141 & 19 & 4\\
\repository{javapoet} & 178 & 6 & 7 & 0 & 2 & 3 & 2 & 0 & 1\\
\repository{sshj} & 2,259 & 48 & 26 & 9 & 9 & 133 & 60 & 24 & 25\\
\repository{RoboBinding} & 12,154 & 893 & 538 & 374 & 233 & 1,188 & 724 & 512 & 262\\
\repository{giraph} & 11,956 & 189 & 118 & 48 & 112 & 154 & 350 & 156 & 261\\
\repository{jeromq} & 2,855 & 64 & 48 & 35 & 45 & 253 & 80 & 58 & 39\\
\repository{jfinal} & 2,760 & 88 & 31 & 9 & 37 & 126 & 111 & 29 & 48\\
\repository{zuul} & 3,386 & 78 & 20 & 60 & 81 & 130 & 134 & 89 & 153\\
\repository{baasbox} & 889 & 15 & 7 & 9 & 2 & 53 & 18 & 8 & 5\\
\repository{spring-data-rest} & 6,814 & 92 & 49 & 58 & 21 & 406 & 106 & 169 & 59\\
\repository{truth} & 8,882 & 139 & 35 & 47 & 33 & 359 & 101 & 240 & 105\\
\repository{rest-assured} & 2,930 & 24 & 18 & 19 & 14 & 171 & 132 & 38 & 54\\
\repository{helios} & 2,577 & 56 & 69 & 34 & 12 & 122 & 44 & 52 & 36\\
\repository{cascading} & 12,483 & 238 & 162 & 91 & 101 & 632 & 309 & 169 & 101\\
\repository{goclipse} & 13,164 & 562 & 249 & 161 & 100 & 854 & 447 & 314 & 135\\
\repository{HikariCP} & 4,181 & 138 & 28 & 7 & 25 & 297 & 74 & 40 & 69\\
\repository{rest.li} & 21,545 & 56 & 139 & 58 & 36 & 463 & 1,495 & 216 & 144\\
\repository{hydra} & 11,423 & 185 & 126 & 84 & 81 & 691 & 333 & 93 & 104\\
\repository{blueflood} & 5,661 & 190 & 48 & 76 & 35 & 460 & 199 & 198 & 106\\
\repository{PocketHub} & 6,181 & 40 & 33 & 55 & 91 & 199 & 99 & 112 & 154\\
\repository{xabber-android} & 7,062 & 400 & 149 & 148 & 77 & 626 & 239 & 159 & 142\\
\repository{morphia} & 28,468 & 691 & 324 & 241 & 293 & 1,140 & 659 & 417 & 381\\
\repository{redisson} & 23,374 & 361 & 176 & 97 & 97 & 745 & 362 & 180 & 173\\
\repository{Activiti} & 30,091 & 705 & 408 & 256 & 253 & 1,586 & 878 & 518 & 319\\
\repository{processing} & 32,286 & 1,355 & 1,236 & 715 & 293 & 1,889 & 1,335 & 3,267 & 401\\
\repository{checkstyle} & 29,281 & 489 & 243 & 202 & 209 & 532 & 206 & 213 & 133\\
\repository{libgdx} & 36,557 & 1,151 & 604 & 398 & 163 & 1,188 & 708 & 373 & 331\\
\repository{cgeo} & 37,178 & 328 & 310 & 242 & 111 & 788 & 858 & 254 & 191\\
\repository{JGroups} & 47,192 & 1,124 & 893 & 436 & 639 & 1,145 & 665 & 579 & 285\\
\hline
\end{tabular}}
\end{table*}

\cref{t:dataset_granularity} shows the number of refactorings detected from each repository.
The second column expresses the number of original refactorings detected, ranging from 178 to 47,192.
The number of CGRs and EPRs detected from granularity level two to granularity level five in each repository is also presented in the table.

\subsection{Preliminary Study on Detection Precision and Efficiency under Granularity Changes}\label{ss:tool_accuracy_performance}

Since changes in commit history granularity can lead to commits containing more changes, potentially affecting the precision and efficiency of refactoring detectors, we conducted a preliminary study to evaluate the precision and efficiency of RefactoringMiner under varying granularity levels.
We applied the techniques introduced in \cref{sec:methodology} to extract CGRs from the \repository{mbassador} repository.
We then manually reviewed the detected CGRs and measured the time required for detection.

\subsubsection{Precision}

The first author manually reviewed all 55 CGRs identified in the repository \repository{mbassador}.
Details regarding the number of commits per CGC at each granularity level can be found in \cref{t:dataset_granularity}.  
Among the 55 CGRs, two were found to be false positives, and we briefly describe each of them below.

The first one is a CGR at $\Level=4$, detected in a CGC squashed by four commits\footnote{\url{https://github.com/bennidi/mbassador/commit/{21385b6,31e6ca1,a47b632,89f4454}}}.
Initially, there are two methods \Method{testRemove1} and \Method{testRemove2()}.
During the code change, \Method{testRemove1()} was renamed to \Method{testRemove2()} with formatting modifications, while the original \Method{testRemove2()} was removed, and a new method \Method{testCompleteRemoval()} was introduced.
RefactoringMiner incorrectly reported a CGR of \Refactoring{Rename Method} from \Method{testRemove1()} to \Method{testCompleteRemoval()}.
While the two methods share some similar logic related to removal functionality, they serve different testing purposes.

The second false positive was found in a CGC at $\Level=5$, consisting of five commits\footnote{\url{https://github.com/bennidi/mbassador/commit/{d08470c,1ed1bb2,f2a61c1,c83d870,32bed9a}}}.  
Initially, class \Class{EventListener2} extended \Class{EventListener1}, and \Class{EventListener3} extended \Class{EventListener2}.  
Both \Class{EventListener2} and \Class{EventListener3} contained a method \Method{handleString()} with an empty body.  
During the code change, the member method \Method{handleString()} in \Class{EventListener2} was removed, and an annotation of that method in class \Class{EventListener3} was removed.
The RefactoringMiner detected an \Refactoring{Extract SuperClass}.
It claims that \Class{EventListener2} is extracted from \Class{EventListener3}, which is incorrect, as the class structure was already in place and no such extraction occurred.

Based on this review, RefactoringMiner achieved a precision of 96.36\% (53/55) in this repository under granularity changes. 
Although this is slightly lower than the 99.6\% precision reported by the tool's authors~\cite{Tsantalis:TSE:2020:RefactoringMiner2.0}, it remains high.
The decrease is acceptable, as squashing code changes from multiple commits into a single commit increases the complexity of the commit.

\begin{table}[tb]\centering
\caption{RefactoringMiner detection time on \repository{mbassador}}\label{t:time_consumed}
{\scriptsize\begin{tabular}{l|ccr} \hline
$\Level$ & \# FGCs/CGCs & Time(s) & Average Time(s)\\
\hline
1&  295&	374.94&	1.27 \\
2&	198&	343.71&	1.74 \\
3&	175&	404.88&	2.31 \\
4&	159&	407.43&	2.56 \\
5&	148&	418.82&	2.83 \\
\hline

\end{tabular}}
\end{table}

\subsubsection{Efficiency}

The time consumed for detecting refactorings in both the original and squashed versions of the \repository{mbassador} repository is summarized in \cref{t:time_consumed}. 
We perform refactoring detection using RefactoringMiner 3.0.4 with its default settings. 
The detection is conducted on each commit in the target repository, executed on a MacBook equipped with an Apple M3 Pro chip and 36 GB of RAM.
The table presents the granularity level, the number of FGCs ($\Level=1$) or CGCs ($\Level>1$) targeted for the detection, the total detection time, and the average detection time per FGC/CGC.
Note that the total detection time includes the process initialization cost of RefactoringMiner per commit since it ran at the commit level.
A statistically significant positive correlation is observed between the granularity level and the average detection time per FGC/CGC (Pearson's $r = 0.986$, $p < 0.05$).

This positive correlation is expected, as squashing code changes from multiple commits into a single commit increases the number of code elements that the detector needs to analyze, thereby increasing the average detection time.

In conclusion, changes in granularity level affect the precision and efficiency of RefactoringMiner, but the impact remains within an acceptable range.

\subsection{\RQ{1}: \RQone}
\subsubsection{Motivation}
This RQ aims to investigate the frequency of CGR occurrences in software development and the necessity of incorporating CGR detection support in refactoring tools.
We explore the frequency of CGRs in open-source repositories with various functionalities at different granularity levels.

\subsubsection{Study Design} \label{ss:rq1_study_design}
The techniques introduced in \cref{subsec:overview,subsec:refactoring_location_tracing} are applied to our dataset to extract squash units, change the granularity of commits, and match the refactoring detection results to find CGRs.

The frequency of CGRs appearing in commit history $C$ of granularity $\Level$, i.e., the ratio of CGR to original refactorings, is presented as follows:
\begin{align*}
  \Frequency^\CGR(H, \Level) &=
    \frac{|\bigcup_{c \in C^\Level(H)}\CGR(c)|}{|\OGR(H)|}
\end{align*}
where $\OGR(H)$ represents the original refactorings detected in $H$.

In this study, we refined the definition of CGR, ensuring that CGRs from different granularity levels are disjoint.
To validate the conclusions from our previous work, which was based on the earlier definition where CGRs from different granularity levels overlapped, we calculated the \textit{accumulated frequency}.
The accumulated Frequency for CGR at granularity level $\Level$ is the cumulative frequency of CGRs at granularity levels up to and including $\Level$.

We calculate the frequencies and the accumulated frequencies of CGRs for all repositories in the dataset at the granularity levels of $\Level \in \{2,3,4,5\}$.

Additionally, we calculate the relative amounts of CGR, EPR, and remaining refactorings based on their detected occurrences for each repository in our dataset.
The relative amount of each type is determined as the proportion of its detected count to the total detected count across all three types.

\subsection{Results and Discussion}\label{ss: rq1_results}

\begin{figure}[tb]\centering
  \includegraphics[width=\linewidth]{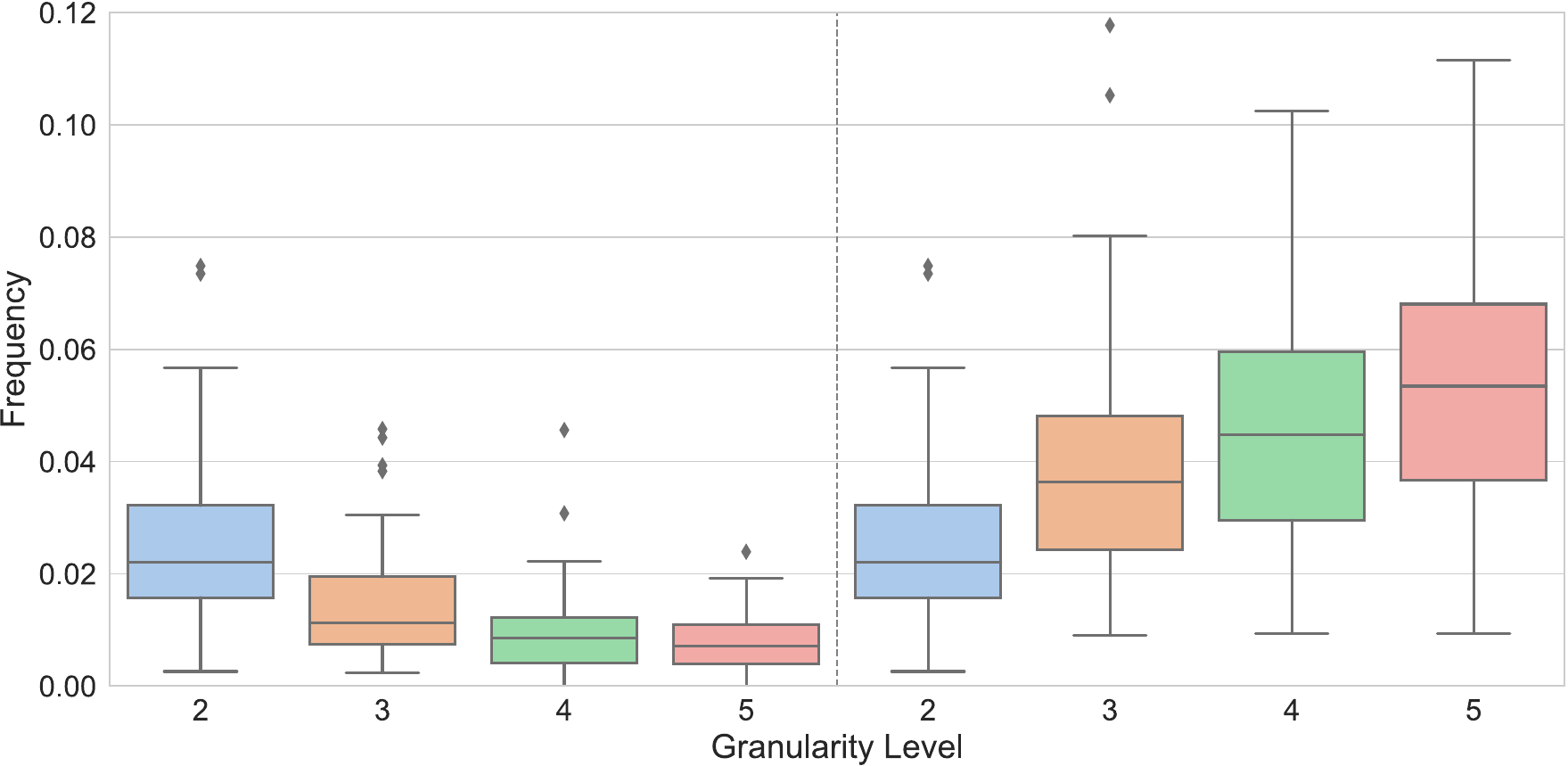}
  \caption{Frequency and accumulated frequency of CGRs.}\label{f:frequency_CGR}
\end{figure}

\begin{figure}[tb]\centering
  \includegraphics[width=\linewidth]{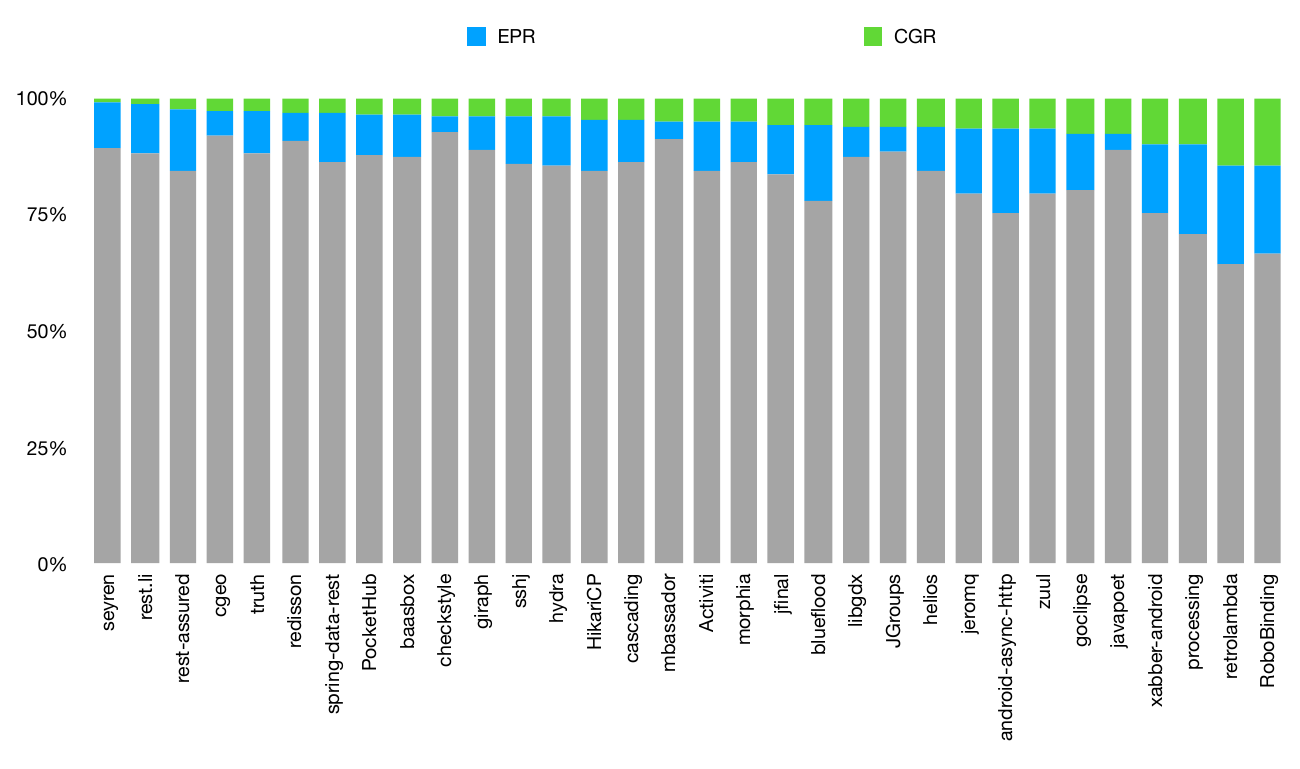}
  \caption{Relative amount of CGRs and EPRs for 32 repositories.}\label{f:relative_amount}
\end{figure}

In \Cref{f:frequency_CGR}, the four boxes on the left represent CGR frequencies, while the four on the right show accumulated CGR frequencies at granularity levels of $\Level \in \{2,3,4,5\}$ across 32 repositories.
The x-axis represents the granularity levels, ranging from $\Level=2$ to $\Level=5$, while the y-axis measures the frequency/accumulated frequency of CGR occurrences.
\Cref{f:relative_amount} depicts the 
relative amount of CGR and EPR against all the refactorings in our dataset.

For the CGR frequencies as shown in \Cref{f:frequency_CGR}, the majority of the boxes for each granularity level indicate that CGR frequencies fall within the range of 0.0 to 0.0566.
The average appearance frequencies of CGRs for all repositories are 0.0256, 0.0158, 0.0101, and 0.0080 when the granularity level is 2, 3, 4, and 5, respectively.

However, certain repositories exhibit notably higher frequencies at specific granularities. 
At granularity levels of 2 and 4, the repositories \repository{retrolambda} and \repository{RoboBinding} show notably higher frequencies.
Specifically, at the granularity level of 2, their frequencies are 0.0749 and 0.0735, respectively, while at the granularity level of 4, they record 0.0456 and 0.0308.
At the granularity level of 3, repositories \repository{android-async-http} and \repository{RoboBinding} show the highest frequencies, at 0.0458 and 0.0443, respectively.

Occurrences of CGRs are a common phenomenon across repositories from various domains and granularity levels. 
As shown in \Cref{f:relative_amount}, although the relative amounts of CGRs vary across repositories, they are all positive, indicating that CGRs are detected in all repositories.
Regarding different granularity levels, except for three repositories where the frequencies are zero at certain levels, all other repositories exhibit consistently positive CGR frequencies across all granularity levels as shown in \Cref{t:dataset_granularity}.

\begin{figure}[tb]\centering
  \includegraphics[width=\linewidth]{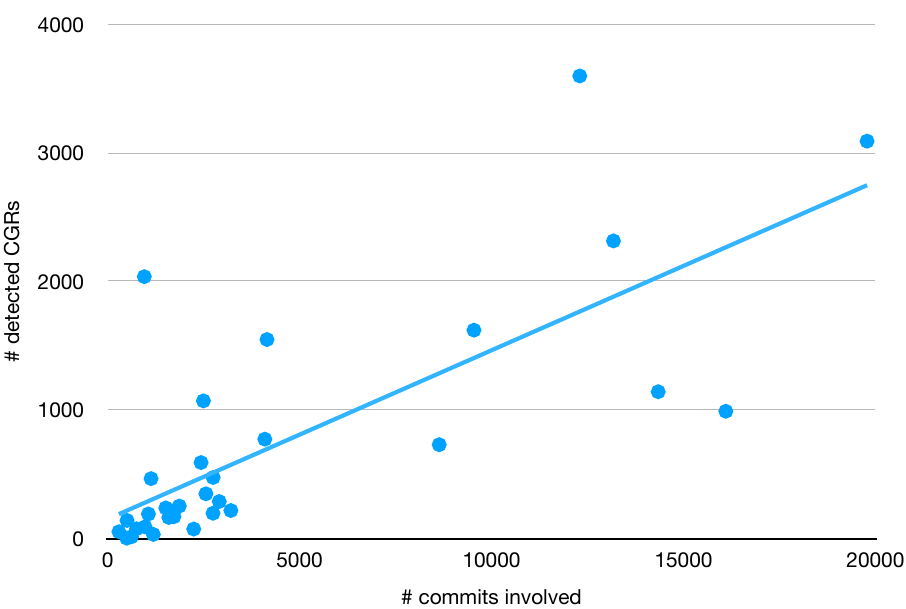}
  \caption{Correlation between number of commits involved in repositories and number of detected CGRs.}\label{f: correlation}
\end{figure}

We also observed a statistically significant positive correlation between the number of commits involved in the straight commit sequences in a repository and the number of detected CGRs, with a Pearson's $r=0.748$ with $p<0.001$.
This relationship is illustrated in the scatter plot in \cref{f: correlation}, where the x-axis is the number of commits involved in the straight commit sequences in a repository, and the y-axis is the number of detected CGRs.
This suggests that the extent of code evolution in a project contributes significantly to the number of CGRs observed.
Projects with more commits typically involve more extensive and iterative changes, which increase the likelihood of CGRs to emerge and be detected.

Nevertheless, the commit count is not the only factor that affects the CGR occurrences across projects.
We observed that \textit{retrolambda} and \textit{seyren}, which have similar commit counts, exhibited notably different numbers of CGRs.
This suggests that additional factors, such as development practices or commitment strategies, may influence the occurrence of CGRs.
Investigating these factors represents a promising direction for future work.

However, the CGR frequency decreases as the granularity level increases.
As the \Cref{f:frequency_CGR} shows, the maximum and median of CGR frequency decrease among the four granularity levels.
Particularly, the median of the granularity level at 5 is 0.0071, the lowest among the measured four granularities.
This is because a CGR is classified at a coarser granularity only when the code change of a single refactoring is distributed across all the consecutive commits in a larger squash unit.
Such occurrences are relatively rare in practical development, making CGRs at the coarsest granularity ($\Level=5$) less frequent.
From another perspective, the cumulative number of CGRs increases progressively with each increment in the granularity level as shown by the right-side four boxes in \Cref{f:frequency_CGR}.
At granularity level 5, the accumulated frequency reaches 0.0571, meaning that for every 100 refactorings performed, approximately 5.71 CGRs are also conducted. 
This exceeds 5\%, emphasizing that the detection of CGRs should not be overlooked.
However, the increase gradually becomes smaller and tends to saturate.
This fact also validates our experimental settings, which set the maximum granularity level at 5, as the frequencies for coarser granularities would be trivial.
Also, it aligns with our previous study's conclusion that when CGRs at different granularities are not disjoint, their frequencies tend to increase as the granularity level rises.

\begin{figure}[tb]\centering
  \includegraphics[width=\linewidth]{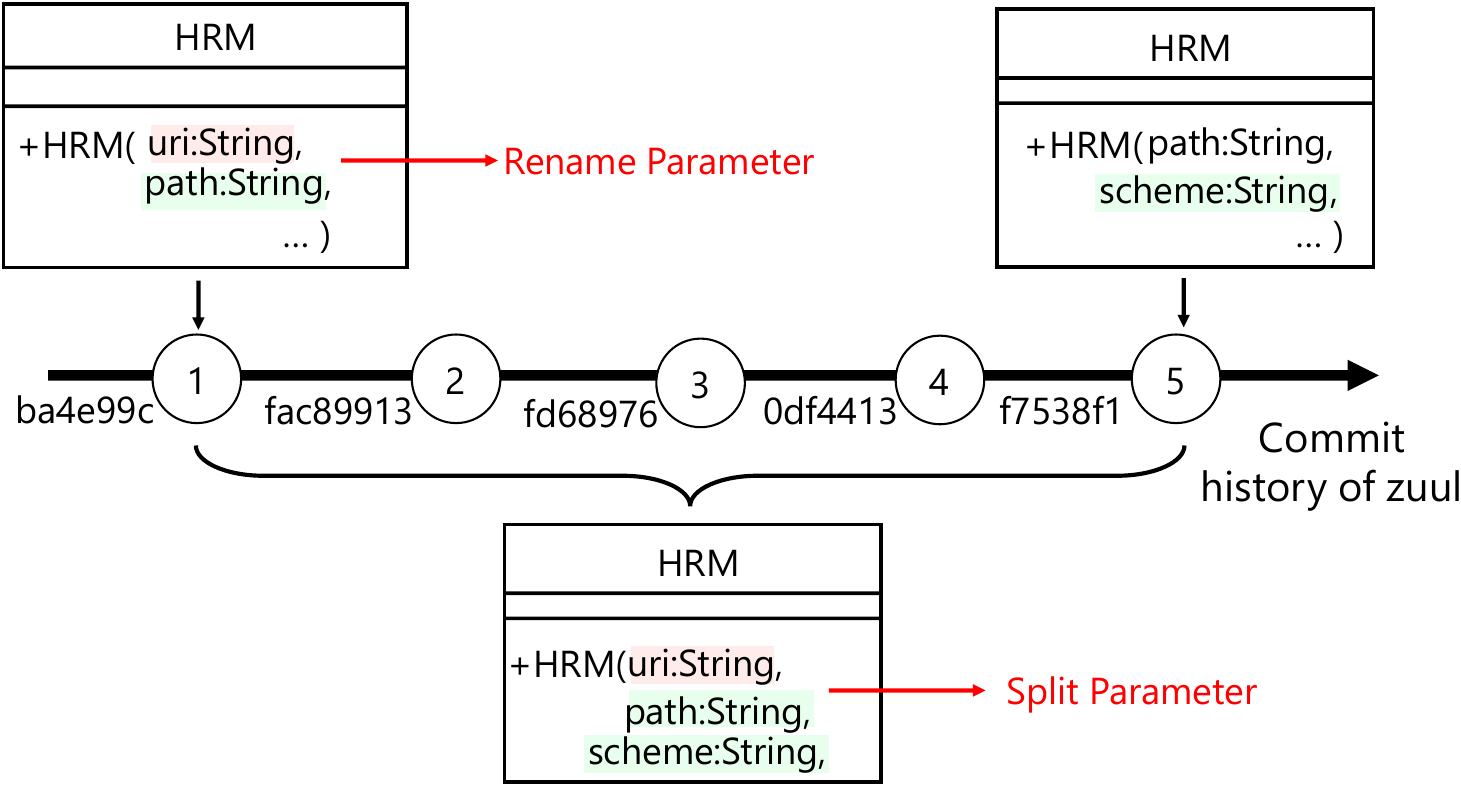}
  \caption{Example of CGR at $\Level = 5$ in \repository{zuul}.}\label{f:granularity_five}
\end{figure}

An example of a coarse-grained refactoring at $\Level = 5$ found in repository \repository{zuul}\footnote{https://github.com/Netflix/zuul/commit/f7538f1} is shown in \cref{f:granularity_five}.
The constructor of class \Class{HttpRequestMessage}~(\Class{HRM} hereinafter) contains several parameters, including \Parameter{uri} of type \Parameter{String} and some other parameters.
In the first proposed commit \CommitId{ba4e99c}, the parameter named \CodeRemoved{uri} of \CodeRemoved{String} is removed, and another parameter \CodeAdded{path} of \CodeAdded{String} is added into the constructor, and the functions related with \CodeRemoved{uri} are replaced by parameter \CodeAdded{path}.
It is detected as a refactoring of \RenameParameter.
In the three commits proposed later, \Class{HRM} is unchanged.
Then, a new parameter \CodeAdded{scheme} of type \CodeAdded{String} is added into the constructor in the fifth proposed commit, and the functions of the parameter \Parameter{path} are replaced by \Parameter{path} and \CodeAdded{scheme}. 
The code change over these five commits is detected as a \SplitParameter, which splits the parameter \CodeRemoved{uri} into parameters \CodeAdded{path} and \CodeAdded{scheme}.

\Conclusion{%
The appearance of coarse-grained refactoring is a common phenomenon in all repositories, and it occurs most frequently at the granularity level of $\Level = 2$.
It is observed that the frequency of CGRs decreases as the granularity level increases.
This trend indicates that the frequency of CGRs decreases as the granularity level increases, and 
On average, according to the number of refactorings conducted, an additional 5.71\% of CGRs are being performed if we see at most the granularity level of $\Level=5$.
}

\subsection{\RQ{2}: \RQtwo}

\subsubsection{Motivation}
This RQ is set to investigate the composition of the detected CGRs in terms of refactoring type.
By answering this RQ, we can provide suggestions about the CGR types that developers of refactoring detectors should prioritize when implementing the CGR-supported detector.
In addition, we also explore the cause of why some types are more frequent than others.

\subsubsection{Study Design}
From the CGRs detected from our dataset, we calculated the appearance ratio of a specific refactoring type at each granularity level.
The appearance ratio expresses the prevalence of one type of CGRs relative to the same type original refactorings.
For a certain refactoring type $t$ in the commit history $C$ at granularity level $\Level$, the appearance ratio can be expressed as follows:
\begin{align*}
    \Appearance^\CGR_t(H, \Level) = \frac%
      {|\{ r \in \bigcup_{ c \in C^\Level(H) } \CGR(c) \mid \Type{r} = t \}|}%
      {|\{ r \in \OGR(H) \mid \Type{r} = t \}|}
\end{align*}
where the granularity $\Level$ of at most 5 was used in this study.
The category of types can refer to Tsantalis et al.~\cite{Tsantalis:TSE:2020:RefactoringMiner2.0}. 

We calculate the appearance ratio of each type of CGR in our dataset.

\subsubsection{Results and Discussions}

\begin{table*}\centering
\caption{Most frequently found CGR types and their appearance ratio.} \label{t:type_rank_CGR}
{\scriptsize\begin{tabular}{cllllll}\hline
       & \multicolumn{4}{c}{Granularity level} \\
  Rank & $\Level = 2$ & $\Level = 3$ & $\Level = 4$ & $\Level = 5$ & cross-granularity \\ \hline
 \multirow{2}{*}{1}
   & \SplitPackage & \SplitPackage & \SplitClass & \MergeClass & \MergeClass \\
   & (29.63\%) & (29.63\%) & (23.44\%) & (18.00\%) & (86.00\%) \\
 \multirow{2}{*}{2}
   & \MergeClass & \MergeClass & \ReplaceAttribute & \SplitClass & \SplitPackage \\
   & (28.00\%) & (26.00\%) & (18.18\%) & (10.94\%) & (81.48\%) \\
 \multirow{2}{*}{3}
   & \ReplaceAttribute & \SplitClass & \SplitPackage & \MergeMethod & \SplitClass \\
   & (27.27\%) & (15.62\%) & (14.81\%) & (8.33\%) & (73.44\%) \\
 \multirow{2}{*}{4}
   & \MergeMethod & \MovePackage & \MergeClass & \SplitPackage & \ReplaceAttribute \\
   & (25.00\%) & (12.88\%) & (14.00\%) & (7.41\%) & (58.18\%) \\
 \multirow{2}{*}{5}
   & \SplitClass & \MoveAndRenameClass & \MergeMethod & \ReplaceAttribute & \MergeMethod \\
   & (23.44\%) & (10.69\%) & (12.50\%) & (7.27\%) & (54.17\%) \\
 \multirow{2}{*}{6}
   & \RenamePackage & \MoveAndRenameMethod & \MoveAndRenameAttribute & \MoveAndRenameMethod & \MovePackage \\
   & (16.26\%) & (9.80\%) & (11.65\%) & (6.22\%) & (41.67\%) \\
 \multirow{2}{*}{7}
   & \MergePackage & \MergeParameter & \MovePackage & \MergePackage & \MoveAndRenameMethod \\
   & (15.15\%) & (9.60\%) & (11.36\%) & (6.06\%) & (39.69\%) \\
 \multirow{2}{*}{8}
   & \MoveAndRenameMethod & \MoveAttribute & \SplitMethod & \MoveAndRenameAttribute & \MoveAndRenameClass \\
   & (14.29\%) & (9.52\%) & (10.61\%) & (6.02\%) & (38.43\%) \\
  \multirow{2}{*}{9}
   & \MoveAndRenameClass & \MergePackage & \MoveAndRenameMethod & \MoveAndRenameClass & \MergePackage \\
   & (13.42\%) & (9.09\%) & (9.38\%) & (5.52\%) & (30.30\%) \\
  \multirow{2}{*}{10}
   & \MovePackage & \MoveMethod & \MoveAndRenameClass & \MovePackage & \MoveAndRenameAttribute \\
   & (12.88\%) & (8.67\%) & (8.81\%) & (4.55\%) & (30.12\%) \\
\hline
\end{tabular}}
\end{table*}

The appearance ratios of each type of CGR across all granularities and the cross-granularity appearance ratios are calculated and ranked.
The top ten are listed in \cref{t:type_rank_CGR}, with their appearance ratio.
In total, 98 distinct CGR types are detected.
The ranking of CGR types varies across granularity levels, indicating that different types of CGRs tend to be applied at specific levels of granularity.

The most frequent CGR types for both granularity levels of 2 and 3 are \SplitPackage with the same appearance ratio: 29.63\%.
For the granularity levels of 4 and 5, the most frequent types are 
\SplitClass and \MergeClass with appearance ratios of 23.44\% and 18.00\%, respectively.

Except for \ReplaceAttribute, all the top three CGR types are associated with splitting or merging operations on packages or classes.
These types of refactorings are inherently complex and require significant modifications, often affecting multiple parts of the code base.
This observation suggests that such complicated, large-scale refactorings typically require incremental modifications spread over multiple commits and appear as CGRs, which highlights the utility of CGRs in facilitating software evolution understanding.
As for \ReplaceAttribute, this refactoring is usually associated with two steps: 1) modifications to the inheritance structure of existing classes, and 2) replacement of an attribute in the subclass with an attribute inherited from the parent class.

\begin{figure*}[tb]\centering
  \includegraphics[scale=0.5]{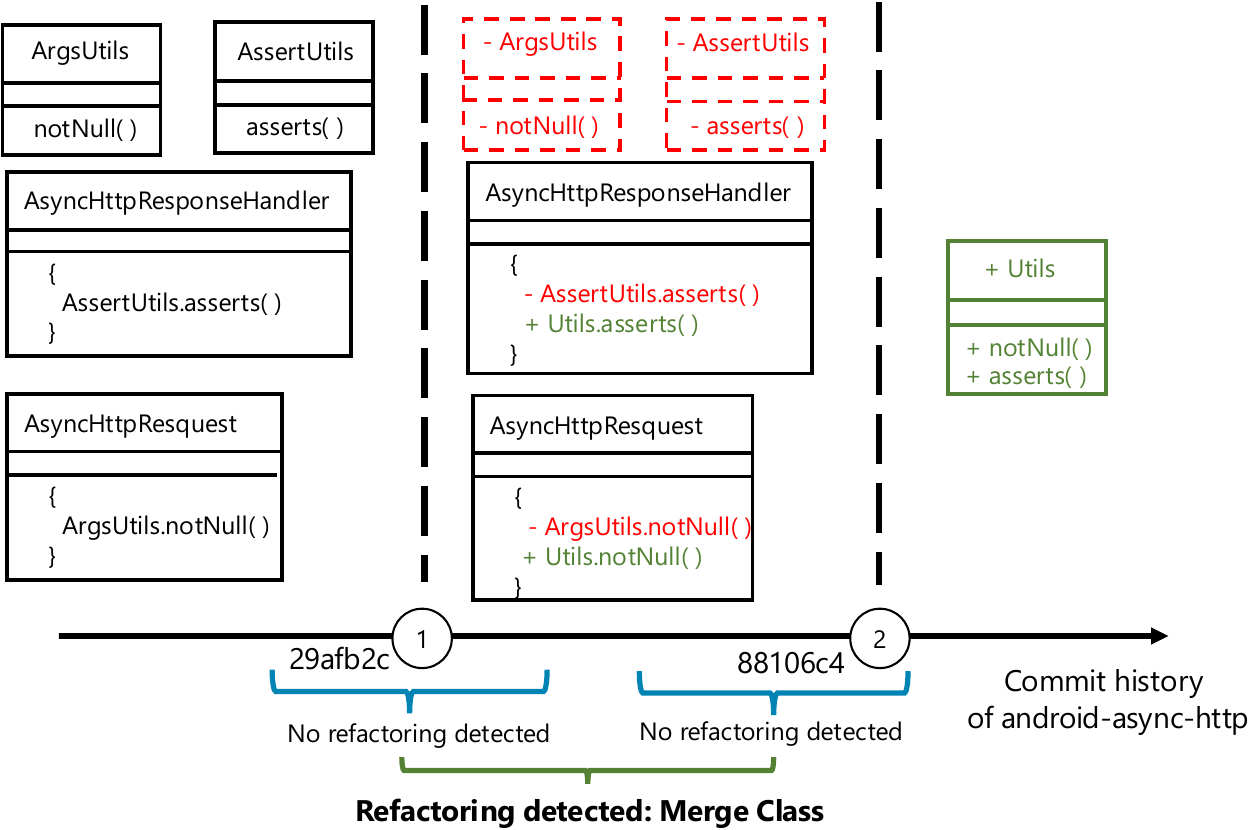}
  \caption{CGR of type \MergeClass found in \repository{android-async-http}.}\label{f:cgr_merge_class}
\end{figure*}

\begin{figure*}[tb]\centering
  \includegraphics[scale=0.5]{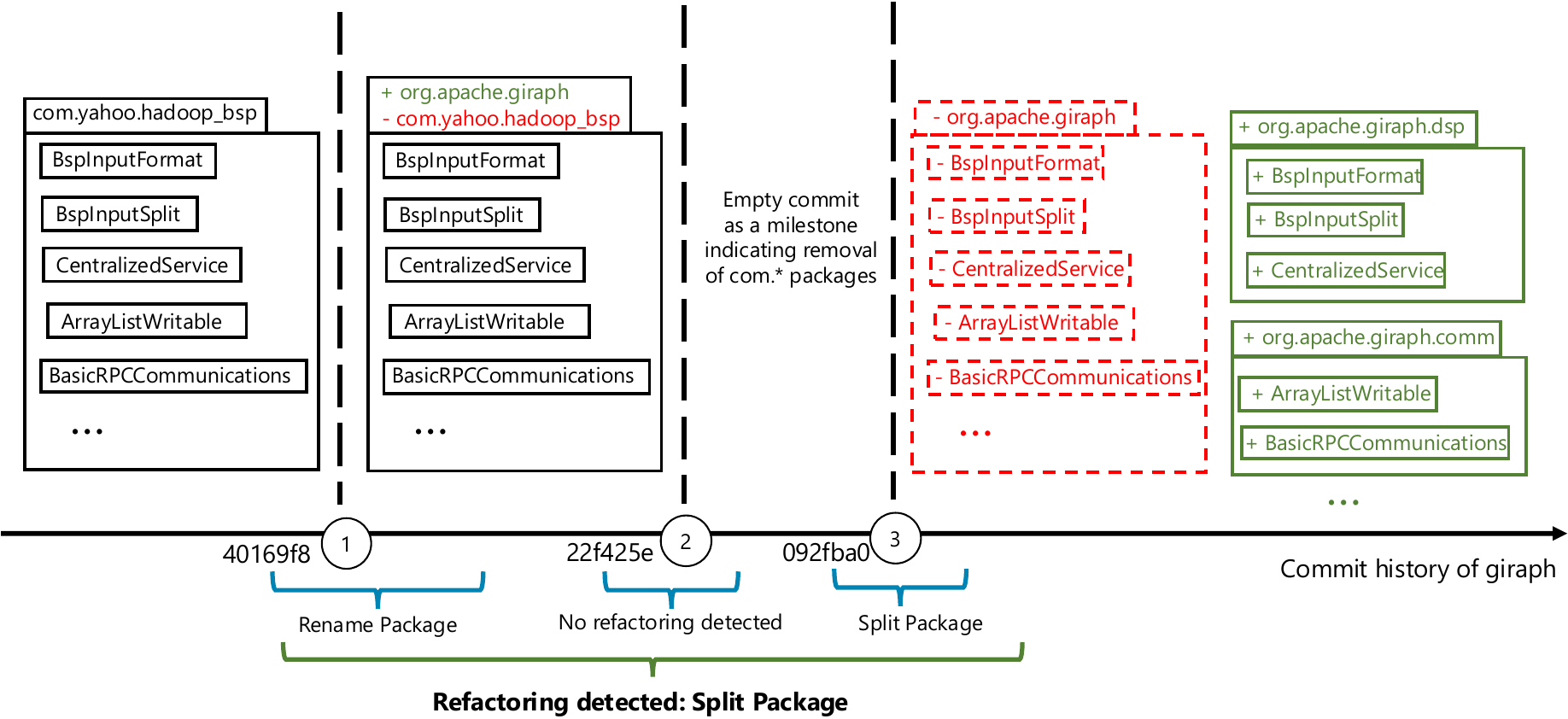}
  \caption{CGR of type \SplitPackage found in \repository{giraph}.}\label{f:cgr_split_package}
\end{figure*}

\begin{figure*}[tb]\centering
  \includegraphics[scale=0.4]{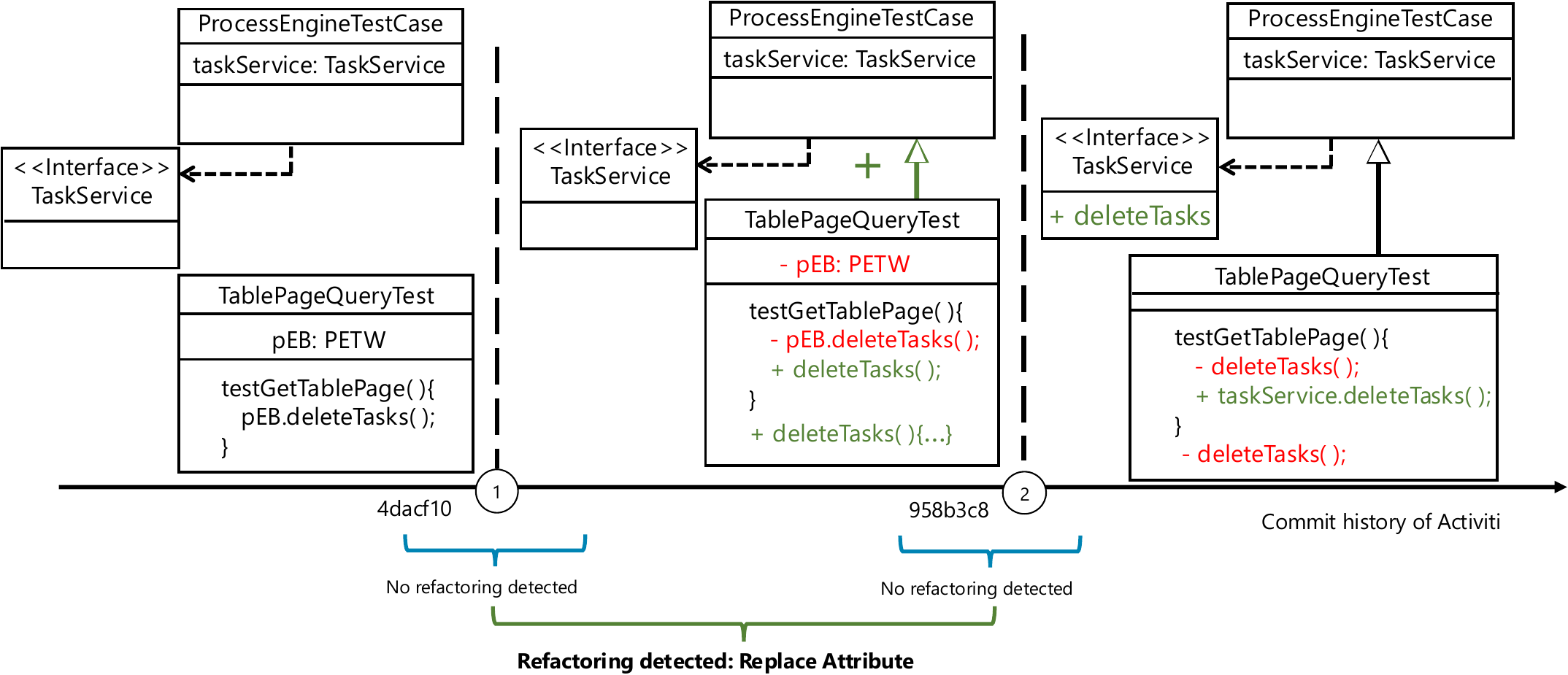}
  \caption{CGR of type \ReplaceAttribute found in \repository{Activiti}.}\label{f:cgr_replace_attribute}
\end{figure*}

We show three examples that contain CGR types of high appearance ratio 
as detailed in the following paragraphs.
The first example is a \MergeClass type CGR, shown in \cref{f:cgr_merge_class}, the second is a \ReplaceAttribute type CGR, shown in \cref{f:cgr_replace_attribute}; and the final example is a \SplitPackage type CGR shown in \cref{f:cgr_split_package}.

The first example is a \MergeClass type CGR detected in repository \textit{android-async-http}\footnote{\url{https://github.com/android-async-http/android-async-http/commit/88106c4}} as shown in \cref{f:cgr_merge_class}.
The \MergeClass is conducted across two commits.
Initially, the \Class{ArgsUtils} and \Class{AssertUtils} classes each contain only a single method: \Method{notNull()} and \Method{asserts()}, respectively.
The method \Method{asserts()} is invoked in methods in class \Class{AsyncHttpResponseHandler}, and the method \Method{notNull()} is invoked in methods in class \Class{AsyncHttpRequest}.
In the first commit, the classes \CodeRemoved{ArgsUtils} and 
\CodeRemoved{AssertUtils} are removed.
The invocation of their methods in classes \Class{AsyncHttpResponseHandler} and \Class{AsyncHttpRequest} are replaced with calls to placeholder methods, \CodeAdded{Utils.asserts()} and \CodeAdded{Utils.notNull()}, defined in an unimplemented \Class{Utils} class that acts as a stub.
The implementation of these stub methods is completed in the second commit \CommitId{88196c4}, where the class \CodeAdded{Utils} is introduced, containing the methods \CodeAdded{asserts()} and \CodeAdded{notNull()} methods with the same method bodies as those in the initial state.
Though no refactoring is detected within either of the commits, a refactoring of \MergeClass is detected across these two commits: the class \Class{AsyncHttpResponseHandler} and \Class{AsyncHttpRequest} is merged into a more consolidated utility class \Class{Utils}.
We deduce that the motivation for splitting the \MergeClass refactoring across two commits is to allow the developer to demonstrate how the new class and methods integrate into the system before finalizing their implementation.
By forwarding the method declarations initially and deferring their implementation, the developer ensures compatibility and minimizes potential disruptions during the transition.

Another example of a \SplitPackage type CGR conducted across three commits is detected in repository \textit{giraph}\footnote{\url{https://github.com/apache/giraph/commit/092fba0}} as illustrated in \cref{f:cgr_split_package}.
In the initial state, the package \Package{com.yahoo.haddop_bsp} contains multiple files.
In the first commit \CommitId{40169f8}, a \RenamePackage refactoring is performed, renaming the package to \CodeAdded{org.apache.giraph}.
The second commit \CommitId{22f425e} is an empty commit with a message indicating that packages starting with \Package{com} are removed.
No refactoring is detected in this commit.
In the third commit \CommitId{092fba0}, a refactoring of \SplitPackage is detected that the package \CodeRemoved{org.apache.giraph} is removed and its files inside are moved to several newly created packages, including \CodeAdded{org.apache.giraph.dsp}, \CodeAdded{org.apache.giraph.comm} and others.
Across these three commits, a refactoring of \SplitPackage is observed.
Different from the one detected in commit \CommitId{092fba0}, the package being split is \Package{com.yahoo.hadoop_bsp}.
Package-related refactorings often have a large-scale impact on the code base.
Instead of directly splitting \Package{com.yahoo.hadoop_bsp} into packages \Package{org.apache.giraph.dsp}, \Package{org.apache.giraph.comm}, and others, the developer applied this operation in two steps: renaming and splitting, and used one empty commit as a milestone to signify the renaming before splitting.
This incremental approach helped manage the transition and ensure the system's stability during the refactoring.

The third example is a \ReplaceAttribute type CGR conducted across two commits detected in repository \textit{Activiti}\footnote{\url{https://github.com/Activiti/Activiti/commit/958b3c8}} as depicted in 
\cref{f:cgr_replace_attribute}.
Prior to the first commit, there are two classes \Class{ProcessEngineTestCase} and  \Class{TablePageQueryTest}.
The \Class{ProcessEngineTestCase} contains an attribute \Attribute{taskService} of type \Class{TaskService}, which is an interface.
The \Class{TablePageQueryTest} has an attribute \Attribute{processEngineBuilder} (\Attribute{pEB}) of type \Class{ProcessEngineTestWatchman} (\Attribute{PETW}), and its method \Method{deleteTasks()} is invoked within the member method \Method{testGetTablePage()}.
In the first commit, the inheritance structure of \Class{TablePageQueryTest} is modified to inherit from the class \Class{ProcessEngineTestCase}, and the attribute \CodeRemoved{pEB} is removed.
The usages of \Attribute{pEB} are replaced by a newly created member method \CodeAdded{deleteTasks()}.
In the second commit, a method declaration of \CodeAdded{deleteTasks()} is added into the interface \Method{TaskService}.
Within class \Class{TablePageQueryTest}, the method \CodeRemoved{deleteTasks()} created in the previous commit is replaced by an invocation of member method \CodeAdded{deleteTasks()} on the attribute \CodeAdded{taskService}, which is inherited from the parent class \Class{ProcessEngineTestCase}.
No refactoring is detected in either of the individual commits.
However, a \ReplaceAttribute refactoring is identified across the two commits.
This refactoring replaces the attribute \Attribute{pEB} in class \Class{TablePageQueryTest} with the inherited attribute \Attribute{taskService} from parent class \Class{ProcessEngineTestCase}.

\Conclusion{%
Refactoring types related to splitting or merging classes and packages, as well as those involving modifications to the inheritance structure, rank highly in CGRs due to their complexity. These refactorings are often implemented incrementally and across multiple commits rather than in a single commit.
}

\subsection{\RQ{3}: \RQthree} \label{ss: RQ3}
\subsubsection{Motivation}
In this RQ, we analyzed the cause of CGRs to gain a comprehensive understanding of how developers introduce CGRs in development.

\subsubsection{Study Design}\label{subsubsec:RQ3_study_design}

We conducted a manual analysis for a sampled set of CGRs.
The first author randomly sampled 110 CGRs from 110 distinct CGC in the dataset.
The sample size of 110 was not based on a specific statistical rationale; rather, it was determined by balancing the required human effort with the goals of the experiment, which is to reveal the typical causes of CGRs.

To reduce potential bias, we ensured that the samples maintain a reasonably balanced distribution across different granularity levels, with a slight emphasis on lower granularity CGRs.
This sampling strategy aligns with our finding in \cref{ss: rq1_results} that lower granularity CGRs tend to occur more frequently.
Among the 110 CGRs, the distribution across granularity levels 2, 3, 4, and 5 was 35, 27, 24, and 24, respectively.
Then, we manually compared and analyzed the conducted changes and refactorings detected, and recognized the cause of why such a change was detected as a CGR.
Based on the recognized causes, we classify the detected CGRs.
Manual analysis can be confirmed in our supplemental package~\cite{dataset}.

\subsubsection{Results and Discussion}
We identified two types of causes for CGRs based on their composition: \textit{Generation} and \textit{Combination}.
Within the sampled 110 CGRs in our dataset, 84 out of 110~(76.4\%) of the CGRs are categorized into \textit{Generation} type, whereas 26 out of 110~(23.6\%) were into \textit{Combination} type.

\Heading{Generation}
This type of CGR is generated from non-refactoring changes.
The example shown in \cref{f:cgr_example} belongs to this type; the \MoveMethod refactoring is generated by two non-refactoring changes: 1) copying the class implementation to a new file and 2) removing the origin class.
Another example is in repository \textit{javapoet}\footnote{\url{https://github.com/square/javapoet/commit/{6a3595c,4ff9adf}}}.
In the parent commit \CommitId{6a3595c}, the attribute \CodeAdded{body} was defined, and the method call \CodeRemoved{methodWriter.write()} was removed.
In child commit \CommitId{4ff9adf}, the developer added method invocation \CodeAdded{body.write()}. 
In the CGC, the above code changes were detected as \RenameVariableWithAttribute; the variable \CodeRemoved{methodWriter} was renamed to attribute \CodeAdded{body}.

\begin{figure}[tb]\centering
  \includegraphics[scale=0.4]{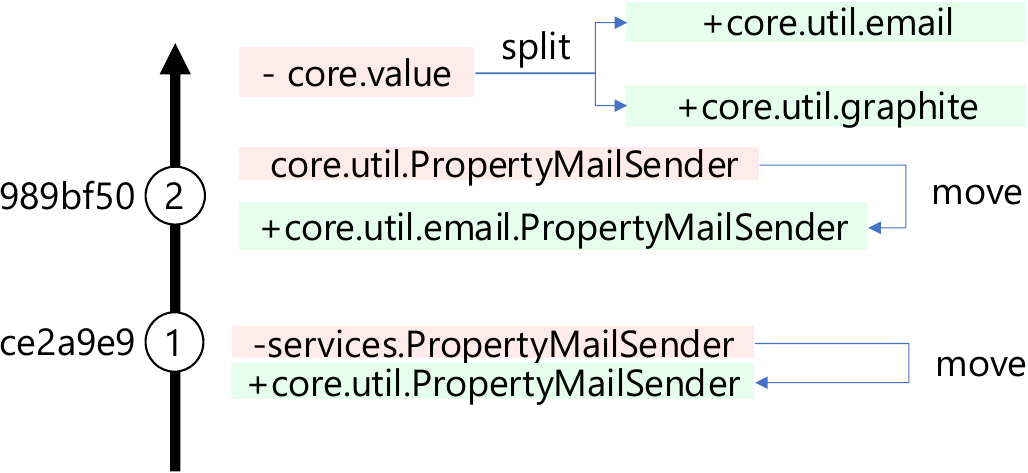}
  \caption{Combination type CGR found in \repository{seyren}.}\label{f:cgr_combination}
\end{figure}

\Heading{Combination}
In contrast with \textit{Generation}, this type is the combined result of multiple refactorings detected in finer-grained commits.
\Cref{f:cgr_combination} shows an example of this type found in repository \repository{seyren}\footnote{\url{https://github.com/scobal/seyren/commit/{ce2a9e9,989bf50}}}.
For clarity, only part of the package hierarchy of the repository is shown in the figure.
In the parent commit \CommitId{ce2a9e9}, the developer moved class \CodeRemoved{PropertyMailSender} under package \CodeRemoved{services} to package \CodeAdded{core.util}, which was detected as refactoring \MoveClass.
In child commit \CommitId{989bf50}, she/he split package \CodeRemoved{core.value} into \CodeAdded{core.util.email} and another one, and then she/he moved class \CodeRemoved{PropertyMailSender} to the package \CodeAdded{core.util.email}, which were detected as \SplitPackage and \MoveClass. 
In terms of result, she/he applied \MergePackage to merge part of the package \CodeRemoved{core.value} and the entire package \CodeRemoved{services} into a new package \CodeAdded{core.util.email}. 

Although we define \textit{Generation} and \textit{Combination} as distinct categories based on whether the CGR is composed of non-refactoring changes or multiple ephemeral refactorings, we acknowledge that borderline cases may exist. 
For example, a CGR identified as \textit{Combination} might have a component refactoring that could be classified as a \textit{Generation} type CGR. 
In such cases, we adopt a conservative classification policy; if any compositional aspect is found, we categorize the CGR as \textit{Combination}.

CGRs of \textit{Generation} type may influence judgments of whether a module is refactored or not.
We note that this type may also occur because of developers' unawareness of refactoring; developers do not realize that the conducted code changes belong to refactoring operations.
Supporting tools to guess developers' manual edits and recognize refactoring activities~\cite{foster2012witchdoctor,ge2014manual} may assist them in development.
Because the \textit{Combination} type may influence type-based refactoring studies, such as investigations on frequently performed refactoring types, researchers may reconsider their results by covering coarse-grained types.

\Conclusion{%
The manual review revealed that 76.4\% of the CGRs were \textit{Generation} type and the rest of 23.6\% were \textit{Combination} type.
\textit{Generation} refers to new refactorings generated by fine-grained non-refactoring changes.
\textit{Combination} is a high-level refactoring combined with more than one refactoring changes.
}

\subsection{\RQ{4}: \RQfour}\label{ss: RQ4}

\subsubsection{Motivation}
We set this RQ to explore whether or not developers will explicitly express their intention of performing CGRs.
The explicitly stated intentions emphasize the necessity of current refactoring detection tools supporting CGR detection, and they are documented in the commit messages~\cite{alomar2019can}.

\subsubsection{Study Design}
The commit messages for the FGCs that are squashed into the 110 CGCs used in answering \RQ{3} were collected and manually reviewed by the first author.
Commit messages were examined to infer the developers' intent, with a focus on identifying the potential relationships among messages from different commits, such as similarity, revert, or other semantic patterns that exist, and suggesting the existence of CGR.
We employed an incremental approach; as we encountered new patterns of intent, we added them to a growing set of characteristics.
Messages that matched one of these characteristics were classified as suggesting the existence of CGRs, while vague or unrelated messages, such as ``update version'' or ``update README'', were excluded.

\subsubsection{Results and Discussions}
The commit messages of FGCs squashed into a CGC are regarded as the commit message of that CGC.
The commit messages of 22 out of 110 CGCs were regarded as explicitly suggesting the existence of CGRs.
We subdivided those commit messages into four types according to their characteristics.

\Heading{Characteristic 1: Same content commit messages (8/110)}
We observed that if the contents of multiple commit messages are the same and contain expressions about performing refactorings, there is a large possibility that CGRs exist.
The same content messages indicate that a single piece of code changes is split into multiple steps, and each step is completed in one commit.
Containing expressions about performing refactorings indicates refactorings are split, which fits the definition of CGRs.
Eight CGCs among the investigated 110 CGCs are found to contain such type commit messages, and we show two of them below.

The first example is three commits extracted from the repository \repository{mbassador}.
The two proposed earlier commits have the same commit messages\footnote{\url{https://github.com/bennidi/mbassador/commit/{014f22d,2ae0e5f}}}: \CommitMessage{introduced specialized subscriptions for better performance and customization options}, which suggests that some subscription-function-related classes are introduced.
The commit message for the remaining commit\footnote{\url{https://github.com/bennidi/mbassador/commit/9ce3ceb}} is \CommitMessage{refactorings. abstract base class. repackaging}, which indicates refactorings about repackaging an abstract class are conducted.
In the CGC squashed by these three FGCs, four \MoveClass CGRs and four \ChangeAccessModifier CGRs are detected.
The cause for the detected CGRs is the same; originally, the classes being moved were in the same file.
In the following commits, the implementations of those classes are copied to different newly created files, and the access modifiers for those classes are changed from private to public. 
And finally, the file which contains their original implementations is removed.
To be more specific, we use two examples to explain.
Originally, the file \File{Mbassador.java} contains two private abstract classes: \Class{Subscription} and \Class{FilteredSubscription}.
In the firstly proposed commit \CommitId{014f22d}, the implementation of class \Class{Subscription} is copied from file \File{MBassador.java} to a new file \File{Subscription.java}, and its class access modifier is changed from private to public.
Then, similarly, in the commit \CommitId{2ae0e5f} proposed later, the implementation of the class \Class{FilteredSubscription} is copied from the same file \File{Mbassador.java} to a new file \File{FilteredSubscription.java} with the change of class access modifier from private to public.
In the last proposed commit \CommitId{9ce3ceb}, the file \File{Mbassador.java} is removed.
The addition of implementation of class \Class{Subscription} in file \File{Subscription.java} and removing of \Class{Subscription} in file \File{Mbassador.java} is regarded as a \MoveClass CGR of $\Level=3$, moving class from \File{Mbassador.java} to \File{Subscription.java}.
Similarly, for the change of class \Class{FilteredSubscription}, it is detected as a \MoveClass refactoring at $\Level=3$.

The second example is found in the repository \repository{RoboBinding}\footnote{\url{https://github.com/RoboBinding/RoboBinding/commit/{e16b566,7d75f31}}}.
The commit messages of the two FGCs have the same content: \CommitMessage{naming improvement to GroupedPropertyViewAttribute}, which suggests the code changes in the commits are renaming operations.
The CGC squashed by these two fine-grained ones contains a \RenameClass CGR of $\Level=2$.
In the first proposed commit, the class \Class{GroupedPropertyAttributeImpl} is removed.
In the later proposed commit, a new class with the same implementation as the removed class is added, and its name is changed to \Class{GroupedAttributeDetailsImpl}.
The only change is the name of the class, and through the commit message, we deduced that the intention of the developers proposing these two commits is to perform a \RenameClass refactoring.

We believe that the existence of this type of CGR indicates that developers split complicated refactorings into steps and perform each step in a single commit to avoid the risk of affecting users.
The consecutive same content commit messages also suggest that the refactoring is not completed in one commit.

\Heading{Characteristic 2: Explicit expression about not finished refactoring (2/110)}
Commit messages of this type are a signal that some refactoring operations are split into several commits.
Two of the 110 CGCs we investigated are found to contain this type, and we show one of them below.
In two commits in the repository \repository{zuul}\footnote{\url{https://github.com/Netflix/zuul/commit/{fac8991,ba4e99c}}}, the commit message of the first commit is \CommitMessage{Further refactoring to use the new filter and context interfaces. Still not complete}.
The keywords ``not complete'' suggest the code change in that commit is incomplete.
The commit next to it but proposed later has a commit message \CommitMessage{And further refactoring to use the new filter and context interfaces}, that suggests this commit is to continue the incomplete refactoring.
The code changes in these two commits also prove our deduction.
In the first commit, the class \Class{HttpServletResponseWrapper} in the file \File{HttpServletResponseWrapper.java} under the path \File{http/} is removed.
In the later proposed commit, the same implementation of the class \Class{HttpServletResponseWrapper} is added in the same name file but under the path \File{servlet/}.
The overall change is a \MoveClass refactoring to move the class file \File{HttpServletResponseWrapper.java} from path \File{http/} to \File{servlet/}.
This type of commit message explicitly points out the existence of CGRs.

\Heading{Characteristic 3: Clearly states the relationship between commits (6/110)}
If the commit messages of multiple commits show some relationships, then CGRs may exist in the squashed commit of these commits.
We found six CGCs contained this type of commit messages and one example is explained below.

In the two consecutive commits \CommitId{fedd975} and \CommitId{8eb25de} in the repository \repository{zuul}\footnote{\url{https://github.com/Netflix/zuul/commit/{fedd975,8eb25de}}}.
The commit message of the first commit \CommitId{fedd975} is \CommitMessage{Revert ``- Added a notifyUsage() template method on FilterProcessor, so that the filter metrics implementation can be overridden''}.
The commit \CommitId{8eb25de} proposed later has a commit message that says \CommitMessage{Reverted the part of the previous commit, as changing from the Singleton pattern for FilterProcessor impacted too many filters that depended on getting access to it}, which reveals that the code change in this commit reverts part of the code change in the previous commit.
The relationship between the two commits is that the latter reverts part of the change made by the previous one.

The code change of the first commit \CommitId{fedd975} reverts the code change of a previous commit \CommitId{7379aa8}.
The commit \CommitId{7379aa8} extracts some code into a method $\Method{notifyUsage()}$, and commit \CommitId{fedd975} removed that method and inlined the method body.
While the second commit \CommitId{8eb25de} extracts the same code into a method $\Method{notify()}$ in a newly created class.

The overall code change of the two commits \CommitId{fedd975} and \CommitId{8eb25de} can be summarized and detected as a \MoveAndRenameMethod CGR, which moves the method $\Method{notifyUsage()}$ to a newly created class and rename it to $\Method{notify()}$.
If we look from commit \CommitId{7379aa8}, a coarser granularity \ExtractMethod CGR is applied to extract the method $\Method{notify()}$, and the \MoveAndRenameMethod CGR is an intermediate step of this refactoring\footnote{
  As indicated by the commit messages, these changes actually correspond to 1) reverting the refactoring applied at one earlier commit (the parent of \CommitId{fedd975}) and 2) conducting refactoring in a different way.
  This combination of reverting and reapplying refactoring could be recognized, at a coarse-grained level, as a distinct type of refactoring.
}.

The relationship in the commit messages indicates that some dependent code changes are performed over multiple commits.
The code change may also be refactorings.
As a result, this characteristic signals the existence of CGRs.

\Heading{Characteristic 4: Existence of connections among commits (6/110)}
If the commit messages of multiple consecutive commits show some connections, then it is possible that CGRs exist.
The difference between this characteristic and the last one is that this type is not so apparent: some commits contain code changes that are performed on the same or related modules, which is not shown explicitly in the commit messages.
We found six out of 110 CGCs containing this type of commit messages, and we show one example below.

For two commits found in the repository \repository{zuul}\footnote{\url{https://github.com/Netflix/zuul/commit/{fc79123,30b8d15}}}, the commit message of the first commit is \CommitMessage{Preserve backward compatible ctor}, and for later proposed commit is \CommitMessage{Remove redundant ctor. Stick to two variants}.
The two commit messages have a connection that both commits are dealing with some constructor (ctor)-related.
From the perspective of code change, the first commit introduced a constructor for class \Class{HttpRequestMessageImpl},  and another constructor for the same class is removed in the latter proposed commit.
The only difference between these two constructors is that the added one has one less parameter than the removed one.
Through these two commits, the developers perform a \RemoveParameter refactoring to remove a parameter in the constructor.

The connection among commits is hard to recognize with not only the commit messages but also the knowledge about the actual code change and original code base.
However, it can signal the existence of CGRs, as shown in the above example.

\Conclusion{%
We found that 20\% (22/110)  of the commit messages explicitly express the existence of CGRs. 
The characteristics of those commit messages are summarized and categorized into four types, where the \textit{Same content commit messages. } is the most frequent characteristic.}

\subsection{\RQ{5}: \RQfive}\label{ss: rq5}
\subsubsection{Motivation}
In answering this RQ, we explore the characteristics of EPRs to gain deeper insights into them through three sub-questions.

\subsubsection{\RQ{5a}\label{sss: rq5a}: How frequently do EPRs appear across the studied repositories?}
\Heading{Study Design}
Similar to CGR, we define the frequency of EPRs at granularity level $\Level$ in commit history $H$ as follows:
\begin{align*}
    \Frequency^{\EPR}(H, \Level) &= \frac%
      {| \bigcup_{c \in C} \EPR^\Level(c) |}%
      {|\OGR(H)|}.
\end{align*}
where $\OGR(C)$ represents the refactorings detected in $C$.
Similar to CGRs, we compute the \textit{accumulated frequency} for EPR at granularity $\Level$ as the cumulative frequency of EPRs at granularity levels up to and including $\Level$.
For all repositories in our dataset, we calculated the frequencies of EPRs detected at granularity levels of $\Level \in \{2,3,4,5\}$.

\begin{figure}[tb]\centering
  \includegraphics[width=\linewidth]{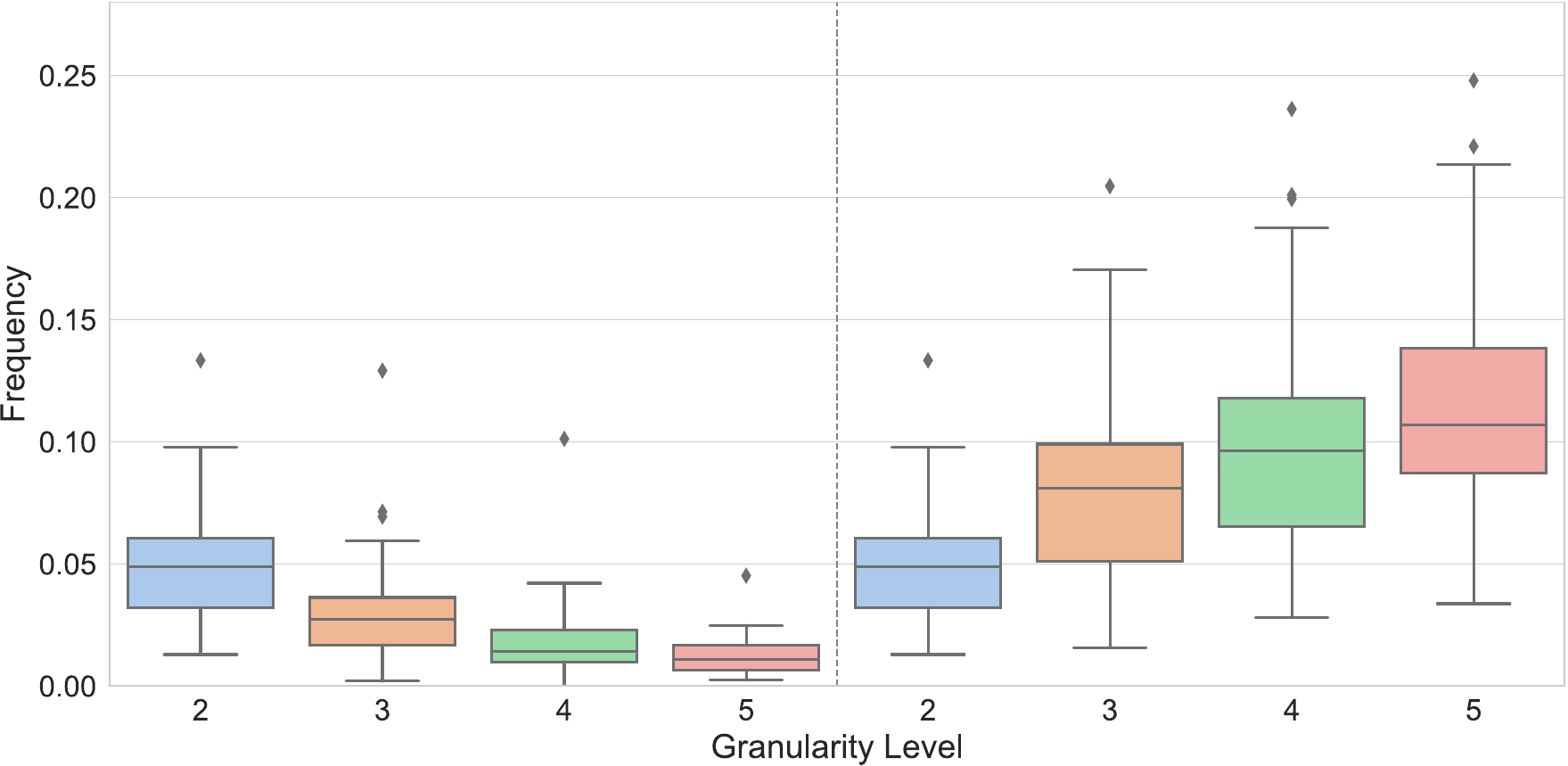}
  \caption{Frequency of EPRs.}\label{f:frequency_EPR}
\end{figure}

\Heading{Results and Discussions}
The frequencies of EPRs detected at granularity levels of $\Level \in \{2,3,4,5\}$ across 32 repositories are depicted in \cref{f:frequency_EPR}.
The x-axis represents the granularity levels, while the y-axis measures the frequency of EPR occurrences.
The boxplots illustrate the distribution of EPR frequencies and accumulate frequencies at each granularity.
The median values of the frequencies for granularity levels of $\Level \in \{2,3,4,5\}$ are 0.0490, 0.0273, 0.0141, and 0.0108, respectively.
The left four boxes indicate a gradual decrease in EPR frequencies as the granularity level increases from 2 to 5.
Additionally, the average value of accumulate frequency reaches 0.0989 when the granularity level is 5, which indicates that per refactoring conducted, there are also 0.0989 EPR be performed.

As introduced in \cref{ss:rq1_study_design} and shown in \cref{f:relative_amount}, EPRs are detected from all repositories, indicating EPR is a common phenomenon.

We conclude that EPR is a common phenomenon among all repositories, and it appears most frequently at the granularity level of 2.

\subsubsection{\RQ{5b}: What are the most common types of EPRs?}

\Heading{Study Design}
The ratio of a specific EPR type $t$ at a granularity level $\Level$ in commit history $H$ can be expressed as follows:
\begin{align*}
  \Appearance^\EPR_t(H, \Level) = \frac%
      {|\{ r \in \bigcup_{c \in H} \EPR^\Level(c) \mid \Type{r} = t \}|}%
      {|\{ r \in \OGR(H) \mid \Type{r} = t \}|}
\end{align*}

We calculated the ratio for each EPR type detected in our dataset.

\begin{table*}\centering
\caption{Most flequently found EPR types and their appearance ratio}\label{t:type_rank_EPR}
{\tabcolsep=4.1pt\scriptsize\begin{tabular}{clllll}\hline
       & \multicolumn{4}{c}{Granularity level} \\
       
  Rank  & $\Level = 2$ & $\Level = 3$ & $\Level = 4$ & $\Level = 5$ & cross-granularity \\ \hline
 \multirow{2}{*}{1}
   & \ModifyParameterAnnotation & \MergeMethod & \RemoveParameterModifier & \SplitVariable & \ModifyParameterAnnotation \\
   & (19.33\%) & (20.83\%) & (17.06\%) & (5.26\%) & (38.66\%) \\
 \multirow{2}{*}{2}
   & \ReplaceAnonymousWithClass & \ModifyParameterAnnotation & \AddParameterModifier & \InlineAttribute & \RenamePackage \\
   & (15.57\%) & (14.29\%) & (12.89\%) & (5.00\%) & (25.12\%) \\
 \multirow{2}{*}{3}
   & \SplitPackage & \ParameterizeAttribute & \RemoveVariableModifier & \SplitClass & \ReplaceAnonymousWithClass \\
   & (14.81\%) & (10.38\%) & (12.49\%) & (4.69\%) & (22.16\%) \\
 \multirow{2}{*}{4}
   & \RenamePackage & \MergeClass & \ReplaceAttribute & \ReplaceVariableWithAttribute & \SplitClass \\
   & (11.82\%) & (8.00\%) & (9.09\%) & (3.06\%) & (21.88\%) \\
 \multirow{2}{*}{5}
   & \SplitClass & \ModifyMethodAnnotation & \SplitClass & \SplitMethod & \MergeMethod \\
   & (10.94\%) & (7.50\%) & (6.25\%) & (3.03\%) & (20.83\%) \\
 \multirow{2}{*}{6}
   & \MergeClass & \ModifyVariableAnnotation & \MovePackage & \MergePackage & \ReplaceAttribute \\
   & (10.00\%) & (6.06\%) & (6.06\%) & (3.03\%) & (20.00\%) \\
 \multirow{2}{*}{7}
   & \MergeVariable & \ReplaceAttribute & \RenamePackage & \SplitParameter & \MovePackage \\
   & (9.68\%) & (5.45\%) & (5.91\%) & (2.88\%) & (19.70\%) \\
 \multirow{2}{*}{8}
   & \MoveAndRenameMethod & \RenamePackage & \ModifyParameterAnnotation & \MoveMethod & \SplitPackage \\
   & (9.38\%) & (5.42\%) & (5.04\%) & (2.28\%) & (18.52\%) \\
 \multirow{2}{*}{9}
   & \MoveAttribute & \LocalizeParameter & \MergeVariable & \MovePackage & \RemoveParameterModifier \\
   & (8.27\%) & (5.35\%) & (4.84\%) & (2.27\%) & (18.43\%) \\
 \multirow{2}{*}{10}
   & \MergeConditional & \InlineAttribute & \SplitPackage & \InlineMethod & \RemoveVariableModifier \\
   & (7.95\%) & (5.00\%) & (3.70\%) & (2.23\%) & (18.31\%) \\
\hline
\end{tabular}}
\end{table*}

\Heading{Results and Discussions}
The top ten EPR types in terms of their appearance ratio for $\Level \in \{2,3,4,5\}$ are listed in \cref{t:type_rank_EPR}.
The highest appearance ratio for the four granularity levels are 
\ModifyParameterAnnotation, \MergeMethod, \RemoveParameterModifier, and \SplitVariable, respectively.
Except for \MergeMethod, \RenamePackage, and \ReplaceAnonymousWithClass, all the other top 3 EPR types across all granularities are operations on small objects such as variables, attributes, and parameters.
Since \RenamePackage targets the name string, it can also be considered an operation on a small object.
This can be explained by the fact that small objects tend to be continuously modified in multiple consecutive commits.

The high appearance ratios of \MergeMethod, \SplitClass, and \SplitPackage can be attributed to the fact that these refactorings have context-sensitive detection criteria; their detection criteria are strict and can be easily disrupted by code changes during squashing.
As a result, they are more likely to be detected as EPRs, which contributes to their high appearance ratios.

\subsubsection{\RQ{5c}: How are EPRs typically introduced in the development process?}

\Heading{Study Design}
The first author manually reviewed and categorized 50 randomly selected EPRs detected in the experiment to investigate the cause of EPRs.
We set the size to 50 based on the balance of our experiment needs and human effort.
The numbers of EPRs at granularity levels 2, 3, 4, and 5 are 22, 16, 7, and 5, respectively.
The distribution aligns with our observation in \RQ{5a}, where the frequency of EPRs decreases as the granularity level increases.
The reviewed result is also included in our supplemental package~\cite{dataset}.

\Heading{Results and Discussions}
We divided the causes of EPRs into three types: \textit{Disruption}, \textit{Absorption}, and \textit{Revert}.

The \textit{Disruption} refers to scenarios where a non-refactoring change violates the match condition of the detection criteria of a refactoring, thereby preventing its identification in the CGC.
For example, in the repository \repository{retrolambda}, commit \CommitId{0971556}\footnote{\url{https://github.com/luontola/retrolambda/commit/0971556}}, renames the class \CodeRemoved{BytecodeFileVisitor} to \CodeAdded{ClasspathVisitor}, which is detected as a \RenameClass refactoring.
However, class \CodeAdded{BytecodeFileVisitor} was newly introduced in its parent commit \CommitId{46b0d84}\footnote{\url{https://github.com/luontola/retrolambda/commit/46b0d84}}.
When analyzing CGC squashed by these two commits, the code change is that class \CodeAdded{ClasspathVisitor} is newly introduced.
The class name \Class{BytecodeFileVisitor} is hidden.
As a result, the refactoring \RenameClass renaming the \Class{BytecodeFileVisitor} to \Class{ClasspathVisitor} is detected as an EPR.

The \textit{Absorption} occurs when multiple refactorings are applied to the same object as part of a refactoring process. 
In the CGC, some of these refactorings are absorbed by the overall process and become undetectable.
For instance, in the repository \repository{mbassador}, commit \CommitId{e7a7628} moved the class \Class{SubscriptionContext} from package \Package{mbassy.dispatch} to \Package{mbassy.subscription} which is detected as a \MoveClass refactoring.
In its parent commit \CommitId{eb50dbc}\footnote{\url{https://github.com/bennidi/mbassador/commit/eb50dbc}}, the same class \CodeAdded{SubscriptionContext} is renamed from \CodeRemoved{MessageContext}, where a \RenameClass refactoring is detected.
When analyzing from the CGC squashed by these two commits, only the \MoveClass refactoring of moving \Class{MessageContext} from \Package{mbassy.dispatch} to \Package{mbassy.subscription} is detected.
The \MoveClass of moving \Class{SubscriptionContext} and the \RenameClass refactoring of renaming \Class{MessageContext} to \Class{SubscriptionContext} are no longer detected, making them EPRs.

The \textit{Revert} refers to a scenario where a refactoring is introduced in one commit but later reverted in a subsequent commit.
An example from the repository \repository{android-async-http} illustrates this.
In commit \CommitId{846c831}\footnote{\url{https://github.com/android-async-http/android-async-http/commit/846c831}}, a \ChangeVariableType refactoring which changed a variable's type from \CodeRemoved{Random} to \CodeAdded{SecureRandom} is detected.
However, in its subsequent commit, another \ChangeVariableType refactoring which reverts this change, restoring the original type is detected.
In the CGC squashed by the above two commits, the variable type remains \textit{SecureRandom}.
As a result, both \ChangeVariableType refactorings are detected as EPRs.

Among the 50 manually reviewed EPR instances, 74\% (38/50) are classified as \textit{Disruption}, 18\% (9/50) as \textit{Absorption}, and 6\% (3/50) as \textit{Revert}. 
Notably, \textit{Revert} EPRs are explicitly mentioned in commit messages, as developers often indicate when a change has been reverted.
In our review, 2 out of 3 Revert-type EPRs were explicitly mentioned in the commit messages.

\Conclusion{%
The median values of frequencies of EPR detected at granularity of 2, 3, 4, and 5 are 0.0490, 0.0273, 0.0141, and 0.0108, respectively.
On average, per refactoring conducted, 0.0989 EPRs will also be performed.
By analyzing the types of the detected EPR, we concluded that refactoring operations on small objects such as variables, attributes, and parameters, and refactorings with context-sensitive detection criteria are often associated with EPRs.
The cause of EPRs is divided into three types: \textit{Disruption}, \textit{Absorption}, and \textit{Revert}.
The \textit{Revert} type EPRs are usually explicitly mentioned in the commit message.
}

\section{Threats to Validity}\label{sec:threats_to_validity}

In this section, we discuss four types of potential threats to the validity of our work as follows:

\subsection{Internal Validity}
One of the possible threats to internal validity is that the repositories selected as datasets are limited in a specific field, which may affect the analysis result.
We mitigated it by selecting 32 repositories from different fields and different companies or organizations.

A second potential threat is the use of \textit{git-blame} for tracking program elements introduced in \cref{ss: matching_scheme}.
Since \textit{git-blame} is inaccurate~\cite{jodavi2022accurate, hasan2024refactoring}, it may affect the correctness of CGR/EPR detection.

Another threat to internal validity is that in \cref{ss: matching_scheme}, when selecting the refactored element to trace in the refactoring detection results, we chose the first code element when multiple elements were present in the detection result based on our observation for sampled refactoring instances that the first element acts as the main role of refactorings.
However, this choice may not be always correct.
The first author manually reviewed all 99 refactoring types detected in this study and found that only 7 might be problematic, as the first element may not always be the primary refactored element. Instances of these types account for only 0.39\% of all detected original refactorings and should not significantly impact the overall findings. Therefore, while this limitation exists, its effect on the validity of our results is minimal.

An additional threat is that in answering \RQ{4}, we analyzed commit messages to assess whether they suggested the existence of CGRs. However, due to the lack of standardized conventions, commit messages are often inconsistent, incomplete, or ambiguous, making them an unreliable proxy for developer intent. 
To mitigate this, we reviewed each message alongside the corresponding code changes and incrementally defined the category of characteristics.
While this approach improves consistency, we acknowledge the inherent subjectivity.
To support better traceability, we recommend that developers explicitly annotate CGRs in commit messages, as noted in the \cref{ss: implications_for_software_practitioners}.

A further threat is that the manual analysis was conducted solely by the first author, which could introduce bias.
To minimize the potential bias, we tried to follow a systematic process for the manual analysis.
We conducted the classification of the cause of CGRs and EPRs through an incremental analysis process; we examined the composition of each instance, identified recurring patterns, and updated the categorization when a new type emerged.
Additionally, to improve transparency and reproducibility, the full classification results are included in the supplemental package~\cite{dataset} so that interested readers can confirm the possibility of misclassified results.

\subsection{External Validity}
One limitation impacting external validity is that although we selected 32 repositories, all of them are programmed in Java.
We cannot claim that the same results can be generalized to other programming languages.

\subsection{Construct Validity}
A possible threat to the construct validity is that we use only one refactoring detector, RefactoringMiner, in this study.
Though it is state-of-art with high accuracy and recall, it may misidentify or miss some refactorings.

\subsection{Conclusion Validity}
A potential challenge to conclusion validity is that in \cref{ss: RQ4}, we deduced the developers' intentions from commit messages for analysis.
Without directly consulting with those developers, our deduction may be inaccurate or incorrect.

Another threat is the limited sample size used in our manual analysis. 
We manually reviewed 110 CGRs for \RQ{3} and \RQ{4}.
For \RQ{3}, we proposed two types of causes for CGR, and we consider this catalog as complete according to the definition of the causes.
Therefore, we do not regard the validity of the classification itself as a threat.
However, due to the limited sample size, the reported ratios of each type may not accurately reflect their true distribution in the dataset.
In answering \RQ{4}, we analyzed commit messages associated with the CGRs to understand developer intentions.
This analysis may also be affected by the limited number of CGRs sampled, and other characteristics may exist.

Similarly, we manually reviewed 50 EPRs for \RQ{5}.
While the three types of causes for EPRs are considered complete based on their definitions, the distribution of each type may be affected by the limited sample size.

In addition, the distribution of the 110 CGRs across granularity levels may also impact our conclusion validity. 
As noted in \cref{subsubsec:RQ3_study_design}, although we maintain a reasonably balanced sample to reduce potential bias, the obtained distribution might be different from the true distribution in the wild.

\section{Implications for Researchers and Practitioners}\label{s: Implication}

Our investigations towards CGRs and EPRs offer several insights that are relevant to refactoring tool builders, empirical software engineering researchers, and software practitioners.
In this section, we discuss how our findings can inform and guide these communities.

\subsection{Implications for Refactoring Tool Builders}

Our investigation of the existence of CGR suggests that the current refactoring tools can be further developed to achieve better performance.
\begin{itemize}
  \item \textbf{Refactoring detector update.}
  Existing modern refactoring detectors typically operate at the granularity of a single commit.
  Our results suggest that refactoring detectors should support detection at multiple levels of granularity, including within a single commit, across multiple commits, and across entire versions, in order to more accurately capture the full scope of refactoring activities.
  For example, RefactoringMiner has a feature for detecting refactorings across a commit range\footnote{\url{https://github.com/tsantalis/RefactoringMiner/tree/f4a2076#with-a-commit-range}}, inspired by our previous work~\cite{chen2022impact}.
  As an advanced detection way, one possible approach to enable CGR detection is as follows: if the match conditions for identifying a refactoring are only partially satisfied in a given commit, the tool could analyze adjacent commits to determine whether the remaining conditions are satisfied in the whole change, including the adjacent commits, thereby enabling detection across commit boundaries.
  \item \textbf{Tool integration.}
  By identifying CGRs and EPRs, integrated development environments (IDEs) and continuous integration (CI) tools can provide more meaningful feedback, such as revealing grouped changes that represent a higher-level design intent or reverted changes that can be regarded as misoperations, thereby improving the user experience during code inspection and review.
  \item \textbf{CGR suggestions.}
  CGRs reflect how developers conduct complicated design changes in practice.
  Tools can identify frequent CGR types and leverage their patterns to recommend multi-step refactorings that align more naturally with developer workflows.
\end{itemize}

\subsection{Implications for Empirical Researchers}

Our findings have methodological and conceptual implications for the empirical study of refactoring and software evolution.
\begin{itemize}
  \item \textbf{Misalignment between change recording and refactoring granularity.}
  Our study highlights a fundamental gap between how code changes are recorded (e.g., as individual commits or pull requests) and the actual granularity of refactorings conducted by developers.
  The commit-level refactoring analyses may systematically underestimate or misrepresent real-world refactoring practices.
  \item \textbf{Dataset construction.}
  The dataset of CGRs and EPRs proposed in this study provides a foundation for more realistic analyses of refactoring in practice.
  Future research can leverage this dataset to evaluate multi-commit refactoring detection approaches, study temporal patterns in code restructuring.
\end{itemize}

\subsection{Implications for Software Practitioners}\label{ss: implications_for_software_practitioners}

Practitioners, including developers and reviewers, can benefit from an improved understanding of CGRs and EPRs:
\begin{itemize}
  \item \textbf{Explicit commit message annotation.}
  Developers can improve the clarity and traceability of CGRs and EPRs by explicitly indicating their presence and intent in their commit messages.
  Such annotations facilitate better understanding during code reviews and inspections.
  \item \textbf{Improved reviewability.}
  Awareness of CGRs and EPRs allows code reviewers to better grasp the overarching refactoring intent, which may be difficult to infer when changes are fragmented across individual commits.
  In addition, such awareness facilitates understanding both individual changes and the broader evolution path of the software.
\end{itemize}

\section{Conclusion and Future Work}\label{sec:conclusion_and_future_work}

In this work, we investigated the impact of refactoring detection on different granularities of commits in 32 open-source Git-based Java repositories.
The granularity change is conducted by integrating the code changes in multiple commits into a single commit.
We named the two types of special refactorings that can only be identified after granularity change as CGR and EPR.
Through the experiments, we observed that the appearance of CGRs is common, and refactoring types related to splitting or merging classes and packages, as well as those involving modifications to the inheritance structure tend to be CGRs.
The cause of CGR is divided into two types according to its composition.
In addition, we also found some developers explicitly suggest the existence of them in the commit message.
The appearance of EPR is also common in development, and refactoring operations on small objects such as variables, attributes, and parameters, and refactorings with context-sensitive detection criteria tend to be associated with EPRs.
We discussed our implications for both researchers and practitioners, which suggests the value of CGRs and EPRs.

For future work, a promising direction is to investigate how the integration path of commits influences the occurrence of CGRs and EPRs. 
Some intermediate states in refactorings may fail to compile or pass CI checks and, therefore, are likely to appear only in pull request workflows where CI is executed on the final merged state.
This can help uncover the strategies developers adopt to perform complex refactorings under different integration workflows.
Additionally, empirical studies that survey how developers perceive CGRs and EPRs and how these affect software engineering tasks such as code review, debugging, and maintenance, represent another valuable direction for future research.
Another complementary direction is to examine the factors that influence the occurrence of CGRs across different projects as we discussed in \cref{ss: rq1_results}.

\section*{Acknowledgments}
This work was partly supported by JST SPRING No.\ JPMJSP2106 and JSPS Grants-in-Aid for Scientific Research Nos.\ JP23K24823, JP25K03102, JP25H01125, JP24H00692, JP21H04877, JP21K18302, and JP21KK0179.

\section*{Declaration of generative AI and AI-assisted technologies in the writing process}
During the preparation of this work the authors used ChatGPT in order to improve readability and language of the work.
After using this tool/service, the authors reviewed and edited the content as needed and take full responsibility for the content of the publication.

\bibliographystyle{elsarticle-num} 
\bibliography{cas-refs}

\begin{thebibliography}{10}
\expandafter\ifx\csname url\endcsname\relax
  \def\url#1{\texttt{#1}}\fi
\expandafter\ifx\csname urlprefix\endcsname\relax\def\urlprefix{URL }\fi
\expandafter\ifx\csname href\endcsname\relax
  \def\href#1#2{#2} \def\path#1{#1}\fi

\bibitem{fowler2018refactoring}
M.~Fowler, Refactoring: {I}mproving the Design of Existing Code, 2nd Edition,
  Addison-Wesley Professional, Boston, MA, 2018.

\bibitem{tsantalis2018accurate}
N.~Tsantalis, M.~Mansouri, L.~M. Eshkevari, D.~Mazinanian, D.~Dig, Accurate and
  efficient refactoring detection in commit history, in: Proceedings of the
  40th International Conference on Software Engineering, 2018, pp. 483--494.

\bibitem{kim2012field}
M.~Kim, T.~Zimmermann, N.~Nagappan, A field study of refactoring challenges and
  benefits, in: Proceedings of the ACM SIGSOFT 20th International Symposium on
  the Foundations of Software Engineering, 2012, pp. 1--11.

\bibitem{kim2014empirical}
M.~Kim, T.~Zimmermann, N.~Nagappan, An empirical study of refactoring
  challenges and benefits at microsoft, IEEE Transactions on Software
  Engineering 40~(7) (2014) 633--649.

\bibitem{bavota2012does}
G.~Bavota, B.~De~Carluccio, A.~De~Lucia, M.~Di~Penta, R.~Oliveto, O.~Strollo,
  When does a refactoring induce bugs? an empirical study, in: Proceedings of
  the IEEE 12th International Working Conference on Source Code Analysis and
  Manipulation, 2012, pp. 104--113.

\bibitem{dig2006automated}
D.~Dig, C.~Comertoglu, D.~Marinov, R.~Johnson, Automated detection of
  refactorings in evolving components, in: Proceedings of the 20th ECOOP
  Object-Oriented Programming, Springer, 2006, pp. 404--428.

\bibitem{kim2010ref}
M.~Kim, M.~Gee, A.~Loh, N.~Rachatasumrit, Ref-finder: a refactoring
  reconstruction tool based on logic query templates, in: Proceedings of the
  18th ACM SIGSOFT International Symposium on Foundations of Software
  Engineering, 2010, pp. 371--372.

\bibitem{prete2010template}
K.~Prete, N.~Rachatasumrit, N.~Sudan, M.~Kim, Template-based reconstruction of
  complex refactorings, in: Proceedings of the 26th IEEE International
  Conference on Software Maintenance, 2010, pp. 1--10.

\bibitem{weissgerber2006identifying}
P.~Wei{\ss}gerber, S.~Diehl, Identifying refactorings from source-code changes,
  in: Processding of the 21st IEEE/ACM International Conference on Automated
  Software Engineering, 2006, pp. 231--240.

\bibitem{silva2017refdiff}
D.~Silva, M.~T. Valente, Ref{D}iff: Detecting refactorings in version
  histories, in: Proceedings of the 14th IEEE/ACM International Conference on
  Mining Software Repositories, 2017, pp. 269--279.

\bibitem{silva2020refdiff}
D.~Silva, J.~P. da~Silva, G.~Santos, R.~Terra, M.~T. Valente, {RefDiff} 2.0: A
  multi-language refactoring detection tool, IEEE Transactions on Software
  Engineering 47~(12) (2020) 2786--2802.

\bibitem{Tsantalis:TSE:2020:RefactoringMiner2.0}
N.~Tsantalis, A.~Ketkar, D.~Dig, {RefactoringMiner} 2.0, IEEE Transactions on
  Software Engineering 48~(3) (2022) 930--950.

\bibitem{ref-det-tools}
L.~Tan, C.~Bockisch, A survey of refactoring detection tools, in: Proceedings
  of the Workshops of the Software Engineering Conference, CEUR-WS, Vol. 2308,
  2019, pp. 100--105.

\bibitem{chen2022impact}
L.~Chen, S.~Hayashi, Impact of change granularity in refactoring detection, in:
  Proceedings of the 30th IEEE/ACM International Conference on Program
  Comprehension, 2022, pp. 565--569.

\bibitem{bibiano2024enhancing}
A.~C. Bibiano, D.~Coutinho, A.~Uch{\^o}a, W.~K. Assun{\c{c}}ao, A.~Garcia,
  R.~de~Mello, T.~E. Colanzi, D.~Ten{\'o}rio, A.~Vasconcelos, B.~Fonseca,
  et~al., Enhancing recommendations of composite refactorings based on the
  practice, in: Proceedings of the 24th IEEE International Conference on Source
  Code Analysis and Manipulation, IEEE, 2024, pp. 83--93.

\bibitem{mongiovi2014making}
M.~Mongiovi, R.~Gheyi, G.~Soares, L.~Teixeira, P.~Borba, Making refactoring
  safer through impact analysis, Science of Computer Programming 93 (2014)
  39--64.

\bibitem{choi2018survey}
E.~Choi, K.~Fujiwara, N.~Yoshida, S.~Hayashi, A survey of refactoring detection
  techniques based on change history analysis, Computer Software 32~(1) (2015)
  47--59, \url{https://arxiv.org/pdf/1808.02320}.

\bibitem{broder1997resemblance}
A.~Z. Broder, On the resemblance and containment of documents, in: Proceedings
  of the Compression and Complexity of SEQUENCES, IEEE, 1997, pp. 21--29.

\bibitem{herzig2013impact}
K.~Herzig, A.~Zeller, The impact of tangled code changes, in: Proceedings of
  the 10th Working Conference on Mining Software Repositories, 2013, pp.
  121--130.

\bibitem{matsuda2015hierarchical}
J.~Matsuda, S.~Hayashi, M.~Saeki, Hierarchical categorization of edit
  operations for separately committing large refactoring results, in:
  Proceedings of the 14th International Workshop on Principles of Software
  Evolution, 2015, pp. 19--27.

\bibitem{sothornprapakorn2018visualizing}
S.~Sothornprapakorn, S.~Hayashi, M.~Saeki, Visualizing a tangled change for
  supporting its decomposition and commit construction, in: Proceedings of the
  42nd IEEE Annual Computer Software and Applications Conference, Vol.~1, 2018,
  pp. 74--79.

\bibitem{ge2014towards}
X.~Ge, S.~Sarkar, E.~Murphy-Hill, Towards refactoring-aware code review, in:
  Proceedings of the 7th International Workshop on Cooperative and Human
  Aspects of Software Engineering, 2014, pp. 99--102.

\bibitem{ge2017refactoring}
X.~Ge, S.~Sarkar, J.~Witschey, E.~Murphy-Hill, Refactoring-aware code review,
  in: Proceedings of the IEEE Symposium on Visual Languages and Human-Centric
  Computing, 2017, pp. 71--79.

\bibitem{alves2014refdistiller}
E.~L. Alves, M.~Song, M.~Kim, Ref{D}istiller: A refactoring aware code review
  tool for inspecting manual refactoring edits, in: Proceedings of the 22nd acm
  sigsoft international symposium on foundations of software engineering, 2014,
  pp. 751--754.

\bibitem{gorg2005detecting}
C.~Gorg, P.~Wei{\ss}gerber, Detecting and visualizing refactorings from
  software archives, in: Proceedings of the 13th International Workshop on
  Program Comprehension, 2005, pp. 205--214.

\bibitem{tsutsumi2016graph}
S.~Tsutsumi, E.~Choi, N.~Yoshida, K.~Inoue, Graph-based approach for detecting
  impure refactoring from version commits, in: Proceedings of the 1st
  International Workshop on Software Refactoring, 2016, pp. 13--16.

\bibitem{sousa2020characterizing}
L.~Sousa, D.~Cedrim, A.~Garcia, W.~Oizumi, A.~C. Bibiano, D.~Oliveira, M.~Kim,
  A.~Oliveira, Characterizing and identifying composite refactorings: Concepts,
  heuristics and patterns, in: Proceedings of the 17th International Conference
  on Mining Software Repositories, 2020, pp. 186--197.

\bibitem{bavota2015experimental}
G.~Bavota, A.~De~Lucia, M.~Di~Penta, R.~Oliveto, F.~Palomba, An experimental
  investigation on the innate relationship between quality and refactoring,
  Journal of Systems and Software 107 (2015) 1--14.

\balance

\bibitem{bibiano2019quantitative}
A.~C. Bibiano, E.~Fernandes, D.~Oliveira, A.~Garcia, M.~Kalinowski, B.~Fonseca,
  R.~Oliveira, A.~Oliveira, D.~Cedrim, A quantitative study on characteristics
  and effect of batch refactoring on code smells, in: Proceedings of the 13th
  ACM/IEEE International Symposium on Empirical Software Engineering and
  Measurement, 2019, pp. 1--11.

\bibitem{cedrim2017understanding}
D.~Cedrim, A.~Garcia, M.~Mongiovi, R.~Gheyi, L.~Sousa, R.~De~Mello, B.~Fonseca,
  M.~Ribeiro, A.~Ch{\'a}vez, Understanding the impact of refactoring on smells:
  A longitudinal study of 23 software projects, in: Proceedings of the 11th
  Joint Meeting on Foundations of Software Engineering, 2017, pp. 465--475.

\bibitem{silva2016we}
D.~Silva, N.~Tsantalis, M.~T. Valente, Why we refactor? confessions of {GitHub}
  contributors, in: Proceedings of the 24th ACM SIGSOFT International Symposium
  on Foundations of Software Engineering, 2016, pp. 858--870.

\bibitem{bibiano2021look}
A.~C. Bibiano, W.~K. Assun{\c{c}}{\~a}o, D.~Coutinho, K.~Santos, V.~Soares,
  R.~Gheyi, A.~Garcia, B.~Fonseca, M.~Ribeiro, D.~Oliveira, et~al., Look ahead!
  revealing complete composite refactorings and their smelliness effects, in:
  Proceedings of the 37th IEEE International Conference on Software Maintenance
  and Evolution, 2021, pp. 298--308.

\bibitem{saika2014kinds}
T.~Saika, E.~Choi, N.~Yoshida, A.~Goto, S.~Haruna, K.~Inoue, What kinds of
  refactorings are co-occurred? an analysis of {Eclipse} usage datasets, in:
  Proceedings of the 6th International Workshop on Empirical Software
  Engineering in Practice, 2014, pp. 31--36.

\bibitem{avgustinov2015tracking}
P.~Avgustinov, A.~I. Baars, A.~S. Henriksen, G.~Lavender, G.~Menzel,
  O.~De~Moor, M.~Schafer, J.~Tibble, Tracking static analysis violations over
  time to capture developer characteristics, in: Proceedings of the 37th
  IEEE/ACM IEEE International Conference on Software Engineering, 2015, pp.
  437--447.

\bibitem{hanam2014finding}
Q.~Hanam, L.~Tan, R.~Holmes, P.~Lam, Finding patterns in static analysis
  alerts: improving actionable alert ranking, in: Proceedings of the 11th
  Working Conference on Mining Software Repositories, 2014, pp. 152--161.

\bibitem{jodavi2022accurate}
M.~Jodavi, N.~Tsantalis, Accurate method and variable tracking in commit
  history, in: Proceedings of the 30th ACM Joint European Software Engineering
  Conference and Symposium on the Foundations of Software Engineering, 2022,
  pp. 183--195.

\bibitem{hasan2024refactoring}
M.~T. Hasan, N.~Tsantalis, P.~Alikhanifard, Refactoring-aware block tracking in
  commit history, IEEE Transactions on Software Engineering 50~(12) (2024)
  3330--3350.

\bibitem{shiba-jssst202211}
S.~Shiba, S.~Hayashi, {Historinc}: A repository transformation tool for
  fine-grained history tracking (in {Japanese}), Computer Software 39~(4)
  (2022) 75--85.

\bibitem{dataset}
L.~Chen, S.~Hayashi, Supplementary package for an empirical study on the impact
  of change granularity in refactoring detection, figshare,
  \url{https://doi.org/10.6084/m9.figshare.29648675} (2025).

\bibitem{foster2012witchdoctor}
S.~R. Foster, W.~G. Griswold, S.~Lerner, {WitchDoctor}: {IDE} support for
  real-time auto-completion of refactorings, in: Proceedings of the 34th
  International Conference on Software Engineering, 2012, pp. 222--232.

\bibitem{ge2014manual}
X.~Ge, E.~Murphy-Hill, Manual refactoring changes with automated refactoring
  validation, in: Proceedings of the 36th International Conference on Software
  Engineering, 2014, pp. 1095--1105.

\bibitem{alomar2019can}
E.~AlOmar, M.~W. Mkaouer, A.~Ouni, Can refactoring be self-affirmed? an
  exploratory study on how developers document their refactoring activities in
  commit messages, in: Proceedings of the 3rd IEEE/ACM International Workshop
  on Refactoring, 2019, pp. 51--58.

\end{thebibliography}

\end{document}